\documentclass[twocolumn]{emulateapj}
\usepackage{natbib}
\usepackage{amsmath}
\usepackage{threeparttable}
\usepackage{subfigure}

\defcitealias{Gonzalez11}{G11}	 	 
\shorttitle{The Rise-Time of Subluminous SNe~Ia}
\shortauthors{Gonz\'alez-Gait\'an S. et al.}

\begin{document}


\title{The Rise-Time of Normal and Subluminous Type Ia Supernovae*}

\thanks{*Based on observations obtained with MegaPrime/MegaCam, a joint project of CFHT and CEA/DAPNIA, at the Canada-France-Hawaii (CFHT) which is operated by the National Research Council (NRC) of Canada, the Institut National des Sciences de l'Univers of the Centre National de la Recherche Scientifique (CNRS) of France, and the University of Hawaii. This work is based in part on data products at the Canadian Astronomy Data Centre as part of the Canada-France-Hawaii Telescope Legacy Survey, a collaborative project of NRC and CNRS.}
\author{S.~Gonz\'{a}lez-Gait\'{a}n$^{1}$\dag}
\email[\dag]{gonzalez@astro.utoronto.ca}
\author{A.~Conley$^{2}$}
\author{F.~B.~Bianco$^{3,4}$}
\author{D.~A.~Howell$^{3,4}$}
\author{M.~Sullivan$^{6}$}
\author{K.~Perrett$^{1,5}$}
\author{R.~Carlberg$^{1}$}
\author{P.~Astier$^{7}$}
\author{D.~Balam$^{9}$}
\author{C.~Balland$^{7,8}$}
\author{S.~Basa$^{10}$}
\author{D.~Fouchez$^{11}$}
\author{N.~Fourmanoit$^{7}$}
\author{M.~L.~Graham$^{3,12}$}
\author{J.~Guy$^{7}$}
\author{D.~Hardin$^{7}$}
\author{I.~M.~Hook.$^{6,13}$}
\author{C.~Lidman$^{14}$}
\author{R.~Pain$^{7}$}
\author{N.~Palanque-Delabrouille$^{15}$}
\author{C.~J.~Pritchet$^{12}$}
\author{N.~Regnault$^{7}$}
\author{J.~Rich$^{15}$}
\author{V.~Ruhlmann-Kleider$^{15}$}

\affiliation{$^{1}$Department of Astronomy and Astrophysics, University of Toronto, 50 St. george Street, Toronto, ON, M5S 3H4, Canada}
\affiliation{$^{2}$Center for Astrophysics and Space Astronomy, University of Colorado, 593 UCB, Boulder, CO, 80309-0593, USA}
\affiliation{$^{3}$Department of Physics, University of California, Santa Barbara, Broida Hall, Mail Code 9530, Santa Barbara, CA 93106-9530, USA}
\affiliation{$^{4}$Las Cumbres Observatory Global Telescope Network, 6740 Cortona Dr., Suite 102, Goleta, CA 93117, USA}
\affiliation{$^{5}$Network Information Operations, DRDC Ottawa, 3701 Carling Avenue, Ottawa, ON, K1A 0Z4, Canada}
\affiliation{$^{6}$Department of Physics (Astrophysics), University of Oxford, DWB, Keble Road, Oxford, OX1 3RH, UK}
\affiliation{$^{7}$LPNHE, Universit\'e Pierre et Marie Curie Paris 6, Universit\'e Paris Diderot Paris 7, CNRS-IN2P3, 4 Place Jussieu, 75252 Paris Cedex 05, France}
\affiliation{$^{8}$Universit\'e Paris 11, Orsay, F-91405, France}
\affiliation{$^{9}$Dominion Astrophysical Observatory, Herzberg Institute of Astrophysics, 5071 West Saanich Road, Victoria, BC, V9E 2E7, Canada}
\affiliation{$^{10}$Laboratoire d'Astrophysique de Marseille, P\^ole de l'\'Etoile Site de Ch\^ateau-Gombert, 38, rue Fr\'ed\'eric Joliot-Curie, 13388 Marseille cedex 13, France}
\affiliation{$^{11}$CPPM, CNRS-IN2P3 and University Aix Marseille II, Case 907, 13288 Marseille cedex 9, France}
\affiliation{$^{12}$Department of Physics \& Astronomy, University of Victoria, PO Box 3055, Stn CSC, Victoria, BC, V8W 
3P6, Canada}
\affiliation{$^{13}$INAF, Osservatorio Astronomico di Roma, via Frascati 33, 00040 Monteporzio (RM), Italy}
\affiliation{$^{14}$Australian Astronomical Observatory, P.O. Box 296, Epping, NSW 1710, Australia}
\affiliation{$^{15}$DSM/IRFU/SPP, CEA-Saclay, F-91191 Gif-sur-Yvette, France}

\begin{abstract}
We calculate the average stretch-corrected rise-time of type Ia supernovae (SNe~Ia) in the Supernova Legacy Survey. We use the aggregate lightcurves of spectroscopic and photometrically identified SNe~Ia to fit the rising part of the lightcurve with a simple quadratic model. We obtain a lightcurve shape corrected, i~.e. stretch-corrected, fiducial rise-time of $17.02^{+0.18}_{-0.28}$(stat) days. The measured rise-time differs from an earlier finding by the SNLS \citep{Conley06} due to the use of different SN~Ia templates. We compare it to nearby samples using the same methods and find no evolution in the early part of the lightcurve of SNe~Ia up to $z=1$. We search for variations among different populations, particularly subluminous objects, by dividing the sample in stretch. Bright and slow decliners ($s>1.0$) have consistent stretch-corrected rise-times compared to fainter and faster decliners ($0.8<s\leq1.0$); they are shorter by $0.57^{+0.47}_{-0.50}$(stat). Subluminous SNe~Ia (here defined as objects with $s\leq0.8$), although less constrained, are also consistent, with a rise-time of $18.03^{+0.81}_{-1.37}$(stat). We study several systematic biases and find that the use of different fiducial templates may affect the average rise-time but not the intrinsic differences between populations. Based on our results, we estimate that subluminous SNe~Ia are powered by $0.05-0.35M_{\odot}$ of $^{56}$Ni synthesized in the explosion. Our conclusions are the same for the single-stretch and two-stretch parameterizations of the lightcurve.

\end{abstract}

\section{Introduction}
Type Ia supernovae (SNe Ia) are believed to be very bright explosions of carbon-oxygen white dwarfs (WDs) in binary systems that undergo thermonuclear runaway. Despite this consensus, the nature of the companion and explosion mechanism are under debate. The enormous energy release of a SN~Ia takes around 2 weeks to peak in the optical. During this early phase of the SN, when the luminosity rises to maximum, fundamental aspects of the explosion, and possibly of the progenitor, are revealed to the observer. The spectroscopic presence of intermediate mass elements (between carbon-oxygen) such as silicon, besides iron-peak elements, suggests incomplete burning from a subsonic deflagration component in the explosion \citep{Filippenko97,Mazzali07}. A subsequent delayed detonation is typically favored to account for the energy output of SNe~Ia \citep{Khokhlov91,Yamaoka92,Woosley94}. 

The radioactive decay of the $^{56}$Ni synthesized in the explosion powers the lightcurve of SNe~Ia \citep{Truran67,Colgate69}. It is thought that the variety in the amount of $^{56}$Ni gives rise to most of the scatter in SN~Ia luminosities. Values of $0.4-0.8M_{\odot}$ of $^{56}$Ni result in ``normal'' SNe~Ia.  ``Subluminous'' or 1991bg-like objects, very faint and fast evolving SNe~Ia, are produced by small amounts of $^{56}$Ni: $0.07-0.17M_{\odot}$ \citep{Filippenko92b,Leibundgut93}. Slowly evolving and very bright SNe, such as 1991T are generated by $0.85-1.1M_{\odot}$ of $^{56}$Ni \citep{Phillips92}. In this framework, the duration from explosion until maximum light, known as the rise-time, is partly determined by the amount of $^{56}$Ni \citep[e.g][]{Arnett82,Shigeyama92,Branch92,Khokhlov93,Hoeflich93,Dominguez01}.

The picture of the early behavior of SNe~Ia has greatly benefited from the advent of improved aspherical and multi-dimensional simulations of delayed detonations \citep[e.g][]{Bravo08,Kasen09,Maeda10,Roepke10}. These authors show that the quantity of $^{56}$Ni synthesized in the explosion that powers the lightcurve depends primarily on a single parameter that can be characterized by the density at which the supposed transition from deflagration to detonation in the delayed detonation framework occurs. If the amount of burning in the deflagration phase is very large (due to, for example, numerous symmetric sparks), the transition density is low and the detonation produces less $^{56}$Ni. Nevertheless, other factors like the metallicity could affect the luminosity to second order \citep[e.g][]{Timmes03,Gallagher08,Howell09,Neill11,Bravo10}. \citet{Hoeflich10}, for example, predict that a dependence of the early lightcurve on the C/O ratio of the progenitor (determined mainly by the progenitor main-sequence mass) potentially affects the slope of the rise behavior at epochs earlier than 5 days before maximum.

Beyond the physics of the explosion, the rise-time study of SNe~Ia can directly provide information on the progenitor system. \citet{Kasen10} suggests that in the single degenerate (SD) scenario, where a WD accretes material from a main-sequence or a red giant, the early phase of the lightcurve can be affected by the interaction of the companion star with the SN shock, particularly at shorter wavelengths. The magnitude depends on the observing angle, as well as on the size and separation of the companion, and can be directly tested with well sampled early rise-time data. In the double detonation of sub-Chandrasekhar mass WDs, \citet{Fink10} predict faster rise-times for brighter and slowly declining SNe~Ia. They also show that models with additional radioactive decay of other short-lived nuclei like $^{52}$Fe and $^{48}$Cr contribute and magnify the early epoch output of SNe~Ia. 

These models suggest that the lightcurve parameterization of SNe~Ia could need additional corrections besides the standard width-luminosity --and color-luminosity as well as the recent host mass-luminosity \citep{Sullivan10,Kelly10}, especially at early epochs. The inclusion of secondary parameters that correlate with luminosity may provide an improved calibration of SNe~Ia as distance indicators for cosmological studies. On the other hand, secondary parameters might not affect the luminosity directly but still be very important to understand the SNe~Ia themselves.

Subluminous SNe~Ia are an extremely faint population with redder colors. Standard scenarios do not reach the very faint explosions of subluminous SNe~Ia and cannot reproduce their Fe- and Si-composition. \citet{Pakmor10} propose a merger system of two WDs of similar masses ($\simeq0.9M_{\odot}$) that is able to account for the low $^{56}$Ni mass of $\simeq0.1M_{\odot}$. These objects are useful to understand the amount of $^{56}$Ni needed to power a lightcurve, and to investigate possible secondary effects. As subluminous SNe~Ia are preferentially found in older, more metal-rich stellar populations \citep{Hamuy96,Howell01a,Gonzalez11}, they also offer an opportunity to study metallicity effects on the amount of $^{56}$Ni produced.

The early epoch study requires good early sampling of SN lightcurves. Modern rolling surveys like the Sloan Digital Sky Survey (SDSS) and the Supernova Legacy Survey (SNLS) have a good cadence and provide very well sampled pre-maximum lightcurve data. Comparing fiducial stretch-corrected rise-times measured from different groups is an extremely difficult task, particularly because of the rather arbitrary definition of a fiducial, or, $s=1$ template. Ideally, the study of several SN~Ia populations, be it at varying redshift or host environment for instance, should be done with the same techniques and templates. Previous studies include the pioneering work by \citet{Riess99b} who found a $B$-band fiducial rise-time of $19.5\pm0.2$ days using combined nearby data $-10$ days before maximum. \citet{Aldering00} find no evidence of evolution in SN~Ia rise-time and \citet[][hereafter C06]{Conley06} confirmed this by calculating the rise-time of the low-$z$ data compared with higher-$z$ from the SNLS ($z<0.9$) finding consistent rise-times, $19.58^{+0.22}_{-0.19}$ for the nearby and $19.10^{+0.18}_{-0.17}$(statistical)$\pm0.2$(systematic) for the SNLS. \citet{Strovink07} found a fiducial rise-time of $17.44\pm0.39$ days with an algorithm that avoids external templates and fits smooth functions to the photometric data directly. Furthermore, they found individual rise-times in a bimodal distribution around $18.81\pm0.36$ days and $16.64\pm0.21$ days. \citet{Hayden10} propose a new parameterization of the lightcurve: they divide it into a rising and a falling component, and find a stretch-corrected rise-time of $17.38\pm0.17$ days for the SDSS, rejecting the bimodal distribution of \citet{Strovink07}. They also find evidence for a faster stretch-corrected rise-time in slow decliners. \citet{Hayden10b} and \citet{Bianco11} go further to constrain the SD models of \citet{Kasen10} and have largely ruled out red giant companions.

None of the previous studies have included subluminous SNe~Ia. In this work, we make use of the large, homogeneously observed, but intrinsically heterogeneous SN~Ia sample obtained by the SNLS, in particular of the photometric group of low-stretch ($s\leq0.8$) objects found in \citet{Gonzalez11}. We investigate the stretch-corrected rise-time behavior of different SN~Ia stretch populations to look for any secondary variations in the lightcurve besides the standard width-luminosity relation. We study the importance of different templates and lightcurve fitting techniques on the stretch-corrected final rise-time calculations. For proper comparison with other samples, we also analyze nearby and SDSS data with our own techniques. The paper is organized as follows: In \S~\ref{theory}  we present the rise-time parameterization used in our study, in \S~\ref{data} we introduce the data we used, its analysis follows in \S~\ref{analysis} and comparisons between subluminous SNe~Ia and other populations, as well as to previous studies are presented in \S~\ref{discussion}. We conclude in \S~\ref{conclusions}.

\section{Rise-time fit}\label{theory}

\subsection{Lightcurve parameterization}\label{lc-paramet}
We follow a similar approach to C06 to measure the common stretch-corrected rise-time of our combined SN~Ia sample data. We call this stretch-corrected rise-time a ``fiducial'' rise-time. Different SNe~Ia are corrected for lightcurve (LC) shape and normalized to peak flux, after expressing all data in the same filter  -- normally $B$ -- via $K$-corrections. For the $K$-corrections, we use the appropriate spectral energy distribution (SED) based on the epoch, SN's redshift and LC parameters such as stretch and color. For restframe epochs (corrected by dividing by $(1+z)$) before explosion, where the flux is generally zero, the $K$-corrections only include the redshift term correction: $-2.5\log(1+z)$ \citep{Nugent02}. Due to the small number of low-stretch SNe~Ia available, our study of the rise-time focuses on the average properties of the aggregate stretch-corrected LCs, instead of individual rise-times.

The LC fitter used throughout this study is a modified version of the SiFTO algorithm \citep{Conley08,Guy10} that uses SED templates to model SN~Ia LCs. The SED sequence has been trained on a large sample of low-$z$ and SNLS data \citep{Conley08} and descends from the spectral sequence of \citet{Hsiao07}. Although the heterogeneous low-$z$ training sample does not include subluminous nor SN2002cx-like objects, the modified fitter also works well for subluminous objects \citep[][hereafter G11]{Gonzalez11}. The LC is parameterized with a single time-axis stretch parameter in the restframe $B$-band, $s_B$. Each observed filter is stretched differently according to a relation between the wavelength and $s_B$ \citep{Conley08}.  The same SEDs are used to $K$-correct all data to the restframe. Observed epochs are transformed to effective dates, $\tau$, by $\tau=(t-t_0)/s (1+z)$, where $t_0$ is the date of maximum flux in the restframe $B$-band and $z$ is the redshift.

The LC fits normally use all available photometry, including the early time. In order not to affect the rise-time analysis, we fit the LC and calculate stretch with only photometry past a certain phase, i.e. $\tau_t<\tau<35$ days. We call this the ``core'' LC. The transition value from rise-time to core LC, $\tau_t$, considered here ranges from $-10$ to $-8$ days (see \S~\ref{transition}). One can also use other LC parameters based uniquely on post-maximum data, such as $\Delta m_{15}$ \citep{Phillips93,Phillips99}.  Alternatively, \citet[][hereafter H10]{Hayden10} suggest a two-stretch approach for dividing the LC fit around the epoch of maximum light $\tau_0$. If $f(t)$ is the flux, we have:

\begin{equation}
f(t)=
\begin{cases}
f_0\,F((t-t_0)/s_r) & \mathrm{if}\, t<t_0 \\
f_0\,F((t-t_0)/s_f) & \mathrm{if}\, t>t_0 \\
\end{cases}
\end{equation}\label{2s-eq}
\noindent
where $f_0$ is the flux at peak (which is the same for both parts of the LC), $F(t)$ is the fiducial ($s=1$) flux of the template, and $s_r$, $s_f$ are the rise and fall stretches, respectively. This means that for the two-stretch technique, the epoch is given by $\tau=(t-t_0)/s_r (1+z)$ for pre-maximum data and by $\tau=(t-t_0)/s_f (1+z)$ for post-maximum data. In this paper, we use both the single- and two-stretch parameterizations to fit the LCs of SNe~Ia. 

\subsection{Rise-time parameterization}
To study the rise-time, we adopt the description used by \citet{Riess99b,Aldering00,Goldhaber01} and C06. If $\tau$ is the LC phase and $t_r$ is the fiducial rise-time of a $s=1$ SN that we want to constrain\footnote{Note that we use $t_r$ for the average fiducial rise-time and $\tau_r$ for the individual rise-time of each SN uncorrected for stretch}, we have for the flux:

\begin{equation}\label{quadrise}
f(\tau)=a(\tau+t_r)^n + b,
\end{equation}
\noindent
for $\tau>t_r$ and $b$ at earlier epochs. Normally, the nuisance parameter $b$ should be zero, but we investigate possible photometric subtraction errors that could systematically shift fluxes up or down (see \S~\ref{beta}). $a$ is the slope of the rise and is another nuisance parameter in the fit. The quadratic form, $n=2$, is motivated by the characteristic expansion of a hot object in the Rayleigh-Jeans tail of the SN~Ia SED at early times in the $B$- and $V$-filters. The flux is proportional to the temperature and to the radius squared, the latter in a homologous expansion ($r\propto v\tau$) depending linearly on the velocity of the expansion: $f\propto r^2 T\propto v^2(\tau+t_r)^2 T$, where $r$ is the radius, $v$ the rise ``speed'' and $T$ the temperature. Since $\Delta T/T$ is almost constant observationally, and $\Delta v/v$ is also close to constant at early epochs, then $f\propto(\tau+t_r)^2$. Although past studies agree that this quadratic model fits the rise-time behavior quite well, small variations among SN~Ia species are possible, not only for the rise-time $t_r$ but also for the power $n$ of the rise. We therefore investigate such variations as well (see \S~\ref{power}).  

\subsection{$^{56}$Ni radioactive decay}
The radiated luminosity of a SN~Ia is powered by the radioactive decay of $^{56}$Ni through $^{56}$Co to $^{56}$Fe \citep{Arnett82,Arnett85,Branch92}. At peak, this energy is still trapped in nickel and is slowly released by its radioactive decay \citep{Nugent95}:

\begin{equation}\label{ni1}
L_{bol}=\gamma \dot{S}(\tau_r)M_{\mbox{Ni}},
\end{equation}
\noindent
where $M_{\mathrm{Ni}}$ is the mass of $^{56}$Ni, $\gamma\sim1.2$ is the ratio of bolometric to radioactivity luminosities for normal SNe~Ia, and $\dot{S}$ is the luminosity from radioactivity per unit nickel mass given by:
\noindent
\begin{equation}\label{ni2}
\dot{S}=\left(6.31\,e^{-\tau_r/8.8}+1.43\,e^{-\tau_r/111}\right)\times10^{43}\,\mbox{erg}\,s^{-1}\,M_{\odot}^{-1}.
\end{equation}

This comes from the characteristic decay times of $^{56}$Ni and $^{56}$Co of 8.8 and 111 days, respectively, with mean energy release per decay of 1.71MeV and 3.76MeV. Therefore, by measuring the individual rise-time $\tau_r$ and bolometric luminosities $L_{bol}$ of SNe~Ia, we can estimate the amount of $^{56}$Ni synthesized in the explosion under the assumed model. Variations in  the amount of $^{56}$Ni mass produced in the explosion will lead to changes in the rise-time for a fixed peak brightness \citep{Contardo00}. 

The value of $\gamma$ is near unity but could in principle vary slightly depending on the $^{56}$Ni mass, i.~e. for large nickel masses, the radioactivity energy released above the photosphere will escape directly instead of remaining trapped, leading to lower values of $\gamma$. In the absence of a model for its variation, we assume $\gamma=1.2\pm0.2$ \citep{Nugent95} and propagate this uncertainty in our calculations.

\section{Data}\label{data}

\subsection{SNLS}
The five-year (2003-2008) Supernova Legacy Survey took advantage of the MegaCam wide-field imager \citep{Boulade03} to observe supernovae, as part of the Canada-France-Hawaii Telescope (CFHT) Deep Synoptic Survey. It used a rolling search technique suited for full coverage of the LC, including early-time, by imaging every 3-4 days (2-3 days in SN restframe) four fields in four different bands $g_Mr_Mi_Mz_M$, which are similar to the filter system $ugriz$ of the SDSS \citep{Smith02}. The SNLS had two independent photometric reconstructions based in Canada and France \citep{Guy10}. We use the French photometry here, as in the SNLS cosmological analyses \citep{Guy10,Conley11}.

\begin{table*}[htbp]
 \centering
 \caption{Nearby objects used in the rise-time calculation. \\
Name (column 1), redshift (column2), single-stretch (column 3), 2-$s$ rise- and fall-stretch (columns 4 and 5), and source (column 6) are presented.}
  \begin{tabular}{|c|c|c|c|c|c|}
    \hline
    \hline
    Name & Redshift & Stretch & Rise-Stretch & Fall-Stretch & Source \\
    \hline
  SN1990n & 0.0034 & 1.07 & 1.18 & 0.96 & \citet{Lira98} \\
  SN1994ae & 0.0043 & 1.05 & 1.09 & 1.03 & \citet{Riess05},\citet{Altavilla04} \\
  SN1997bq & 0.0094 & 0.91 & 1.05 & 0.81 & \citet{Jha06} \\
  SN1998aq & 0.0037 & 0.97 & 1.10 & 0.90 & \citet{Riess05} \\
  SN2000e & 0.0048 & 1.06 & 1.28 & 1.00 & \citet{Valentini03,Tsvetkov06a} \\ 
  SN2001el & 0.0039 & 0.96 & 1.12 & 0.91 & \citet{Krisciunas03} \\
  SN2001v & 0.0150 & 1.09 & 1.22 & 1.05 & \citet{Hicken09a} \\
  SN2002bo & 0.0042 & 0.94 & 1.02 & 0.93 & \citet{Krisciunas04} \\
  SN2003du & 0.0064 & 1.00 & 1.04 & 1.00 & \citet{Anupama05},\citet{Stanishev07},\citet{Leonard05} \\
  SN2004fu & 0.0092 & 0.90 & 0.99 & 0.84 & \citet{Tsvetkov06c,Hicken09a} \\ 
  SN2005cf & 0.0064 & 0.98 & 1.03 & 0.96 & \citet{Pastorello07},\citet{Wang09},\citet{Hicken09a} \\
  SN2005mz & 0.0176 & 0.63 & 0.91 & 0.53 & \citet{Hicken09a}\\
  SN2006cp & 0.0223 & 1.05 & 1.11 & 1.02 & \citet{Hicken09a}\\
  SN2006gz & 0.0237 & 1.22 & 1.30 & 1.19 & \citet{Hicken07} \\
  SN2006le & 0.0174 & 1.08 & 1.10 & 1.05 & \citet{Hicken09a}\\ 
  SN2007af & 0.0055 & 0.94 & 1.07 & 0.91 & \citet{Hicken09a} \\


\hline
  \end{tabular}
  \label{lowz-data}
\end{table*}

C06 used 73 spectroscopically confirmed normal SNe Ia for their rise-time study. Here, we use 30 photometrically identified low-stretch ($s\leq0.8$) SNe~Ia and 221 normal SNe~Ia \citep{Gonzalez11,Perrett11} at $0.1\leq z\leq0.7$ to study the respective stretch-corrected rise-times and compare them. The photometric candidates went through a selection process that ensured proper redshift and stretch estimation. As part of the selection criteria to constrain the stretch, we require here good lightcurve sampling, particularly enough early data. For this, we re-fit for stretch using only data \emph{after} $\tau_t$, to ensure the lightcurve shape (stretch) is not affecting our rise-time estimation. We require at least two data points within $\tau_t<\tau<0$ days of restframe $B$-band maximum. The restframe $B$-band for these objects will be mostly mapped by $r_M$ (and somewhat by $i_M$). The lightcurves must also have a good SiFTO fit with a total reduced chi-squared, $\chi^2_{\nu}$, better than $3.0$. Additionally, we require that the error in the determination of the peak epoch be less than $0.7$ days ($1.2$ days for the two-stretch parameterization, as shown in section~\ref{2s}). For the rise-time region fit, SNe~Ia are required to have data prior to $-10$ days. These cuts reduce the sample to 134 SNe, of which 12 have $s\leq0.8$.

In our dataset, 35\% of the low-$s$ and 74\% of the normal SNe~Ia have a spectroscopic redshift. The effect of having uncertain photometric redshifts can lead to considerable errors propagated in the $K$-corrections as well as in the effective dates $\tau$ of each SN, ultimately affecting the rise-time fit. Throughout this study, photometric redshifts are used in the same way spectroscopic redshifts are, but a proper inclusion of their errors is analyzed in \S~\ref{rise-calc}.

\subsection{Low-$z$ and SDSS}

To test our techniques, we also analyze available low-$z$ rise-time data (see Table~\ref{lowz-data}), as well as the SDSS-II SNe~Ia from \citet{Holtzman08}. We can thus compare our results to low redshift to study any evolutionary effects. We cannot provide a comparison of our low-$s$ stretch-corrected rise-time analysis to nearby ones, as published early data for nearby $s\leq0.8$ are scarce. The cuts applied to the SNLS are also used for the SDSS-II and low-$z$ data, i.~e. at least two data points in restframe $B$-band at $\tau_t<\tau<0$ days, and at least one point before $\tau_t$. 


\section{Analysis}\label{analysis}

\begin{figure*}[htbp]
        \centering
                \includegraphics[width=0.46\linewidth]{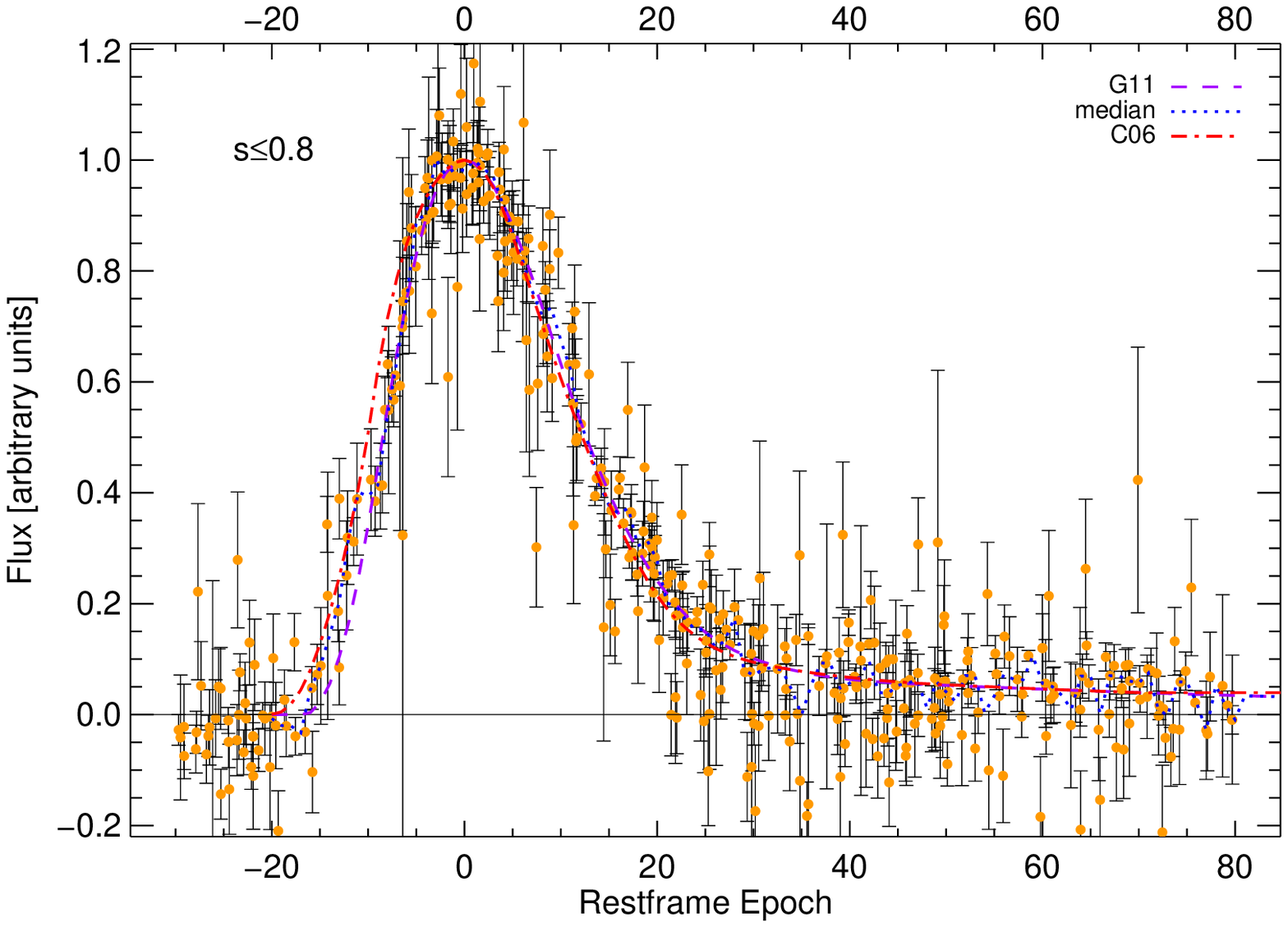}
                \includegraphics[width=0.46\linewidth]{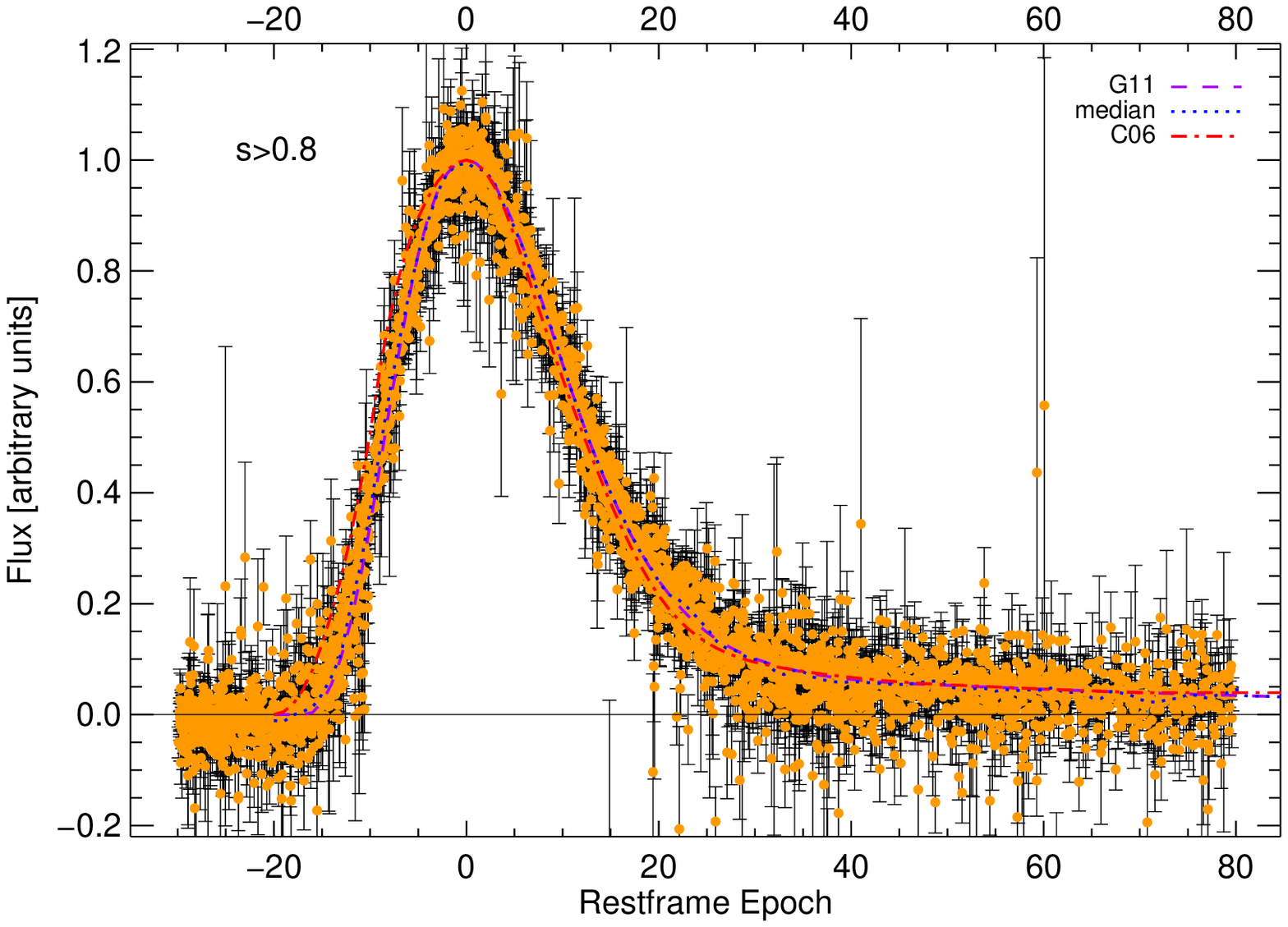}
\caption{Overlaid B-band restframe lightcurves of low- ($s\leq0.8$) and normal-stretch ($s>0.8$) SNe~Ia from the SNLS. These are K-corrected, flux normalized and scaled to $s$ and $(1+z)$. The blue dotted line is the median of the flux points, the dashed violet line is the fiducial G11 template used, whereas the red dot-dashed is the template used in C06.}
        \label{overLC}
\end{figure*}

\begin{figure*}[htbp]
        \centering
                \includegraphics[width=0.46\linewidth]{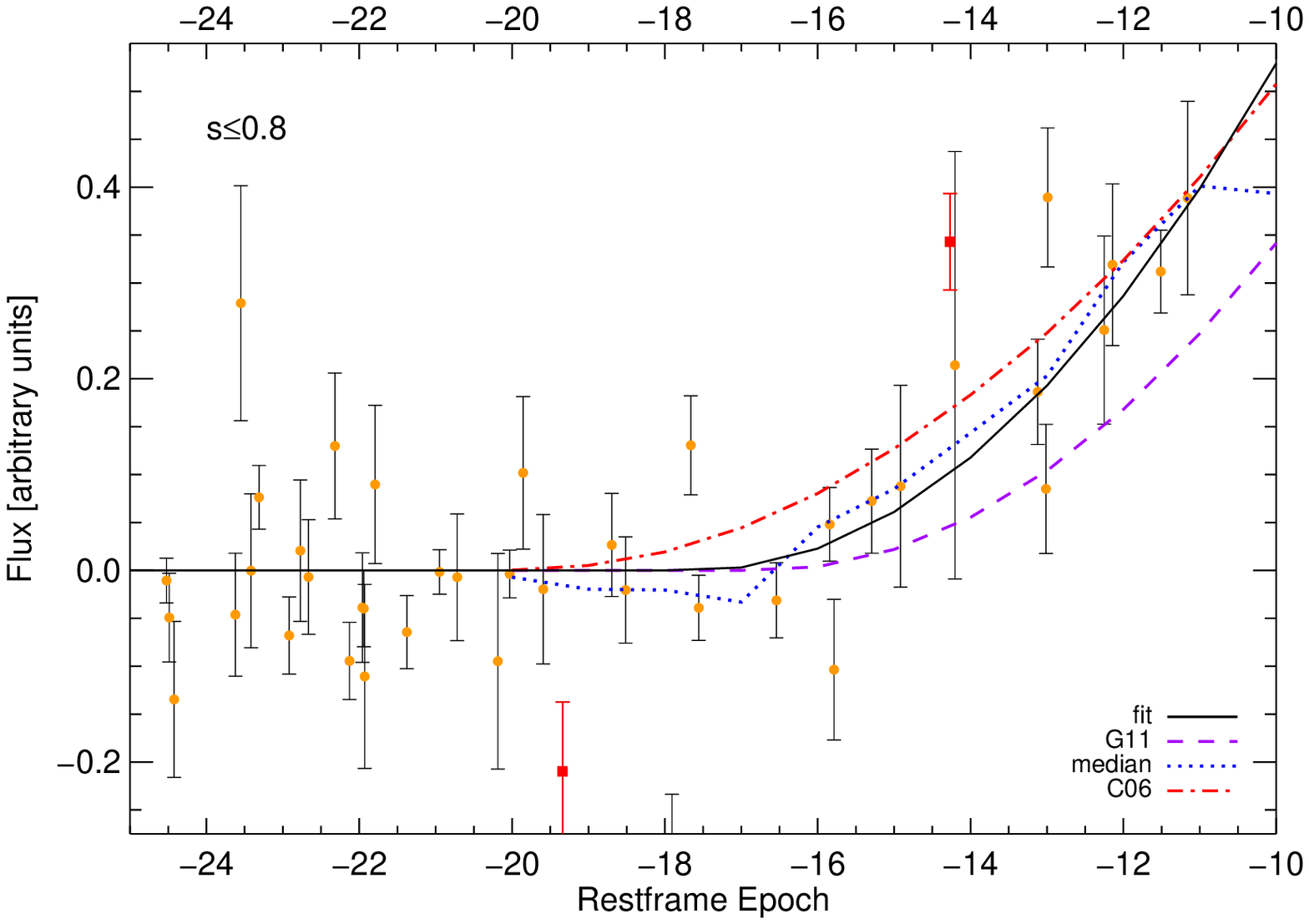}
                \includegraphics[width=0.46\linewidth]{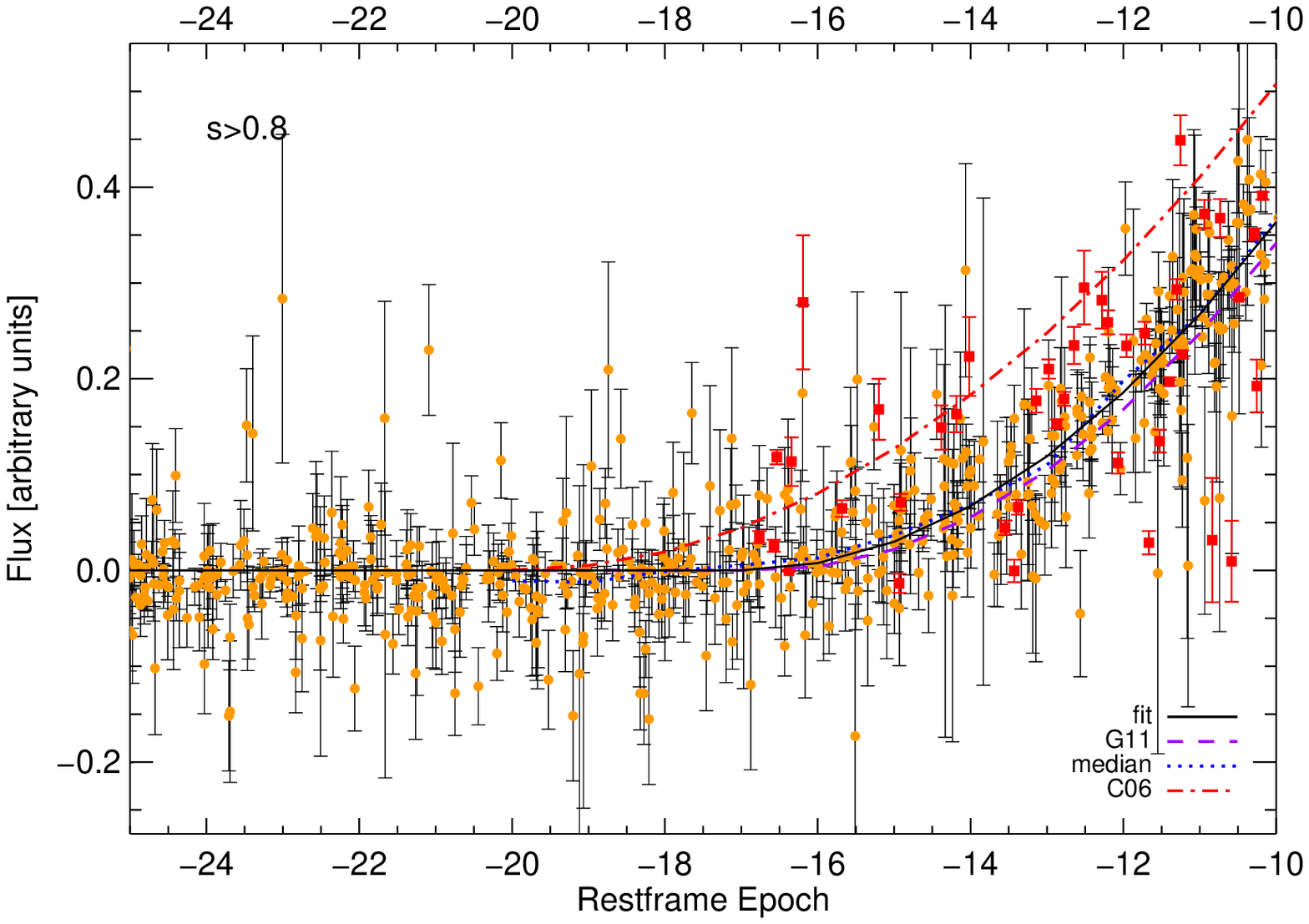}
\caption{Overlaid B-band restframe lightcurves of low- ($s\leq0.8$) and normal-stretch ($s>0.8$) SNe~Ia from the SNLS in the early rise at $\tau_t<-10$ days. Red squares are outliers from the fit ($\sim3.4\sigma$). The solid black line is the best quadratic fit, the blue dotted line is the median of the flux points, the dashed violet line is the fiducial G11 template used, and the red dot-dashed is the C06 template.}
        \label{overLC-early}
\end{figure*}


\subsection{Rise-time calculation}\label{rise-calc}

To make a combined rise-time study of all SNe~Ia in each sample, we process each lightcurve as described in \S~\ref{lc-paramet}. We then take the restframe $B$-fluxes of the low-stretch SNe Ia and overlay them on a single plot, as shown in Figure~\ref{overLC}. For the rise-time region, we use early data, before a characteristic transition epoch $\tau_t=-10$ days to fit it to the model of Equation~\ref{quadrise} (see Figure~\ref{overLC-early}). Minimizing the $\chi^2$ of the best fit model with respect to the data provides the best stretch-corrected rise-time. The choice of the transition epoch used affects the fit, as will be shown in \S~\ref{transition}. We perform an iterative outlier rejection (similar to C06) to discard any data points that strongly deviate from the fit according to Chauvenet's criterion \citep{Taylor97}. For the total sample, this corresponds to $3.4\sigma$ and discards 25 (out of 449) data points, but does not affect the final fit. 

\begin{figure}[htbp]
        \centering
                \includegraphics[width=1.0\linewidth]{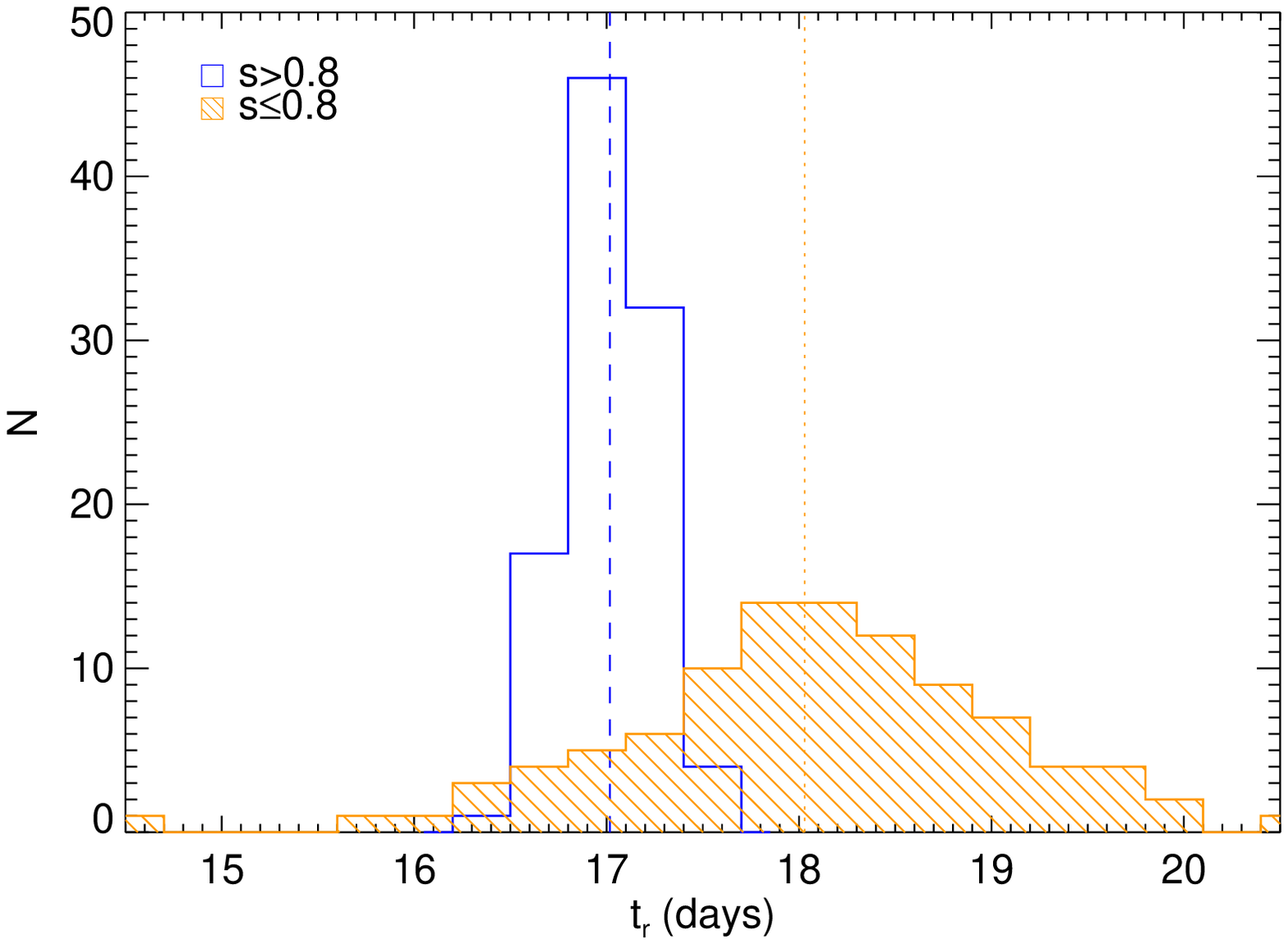}
                \caption{Distribution of 100 MC stretch-corrected rise-time fits to $s\leq0.8$ (filled orange) and $s>0.8$ (empty blue) SNe~Ia. The dashed and dotted lines represent the median for both samples.}
        \label{mcrisehist}
\end{figure}

The early time fit of the aggregate LC, nevertheless, does not take into account uncertainties in the SN fit parameters and the photometric redshift. Errors in the redshift estimate induce small changes in stretch (\citetalias{Gonzalez11}). Additionally, the use of a common SED for the LC fit as well as for the $K$-corrections introduces strong correlations, as do the uncertainties of maximum date and stretch of each SN fit. We handle these by shifting the photometric redshift according to its error (for those SNe~Ia with photo-$z$) in a Monte Carlo (MC) fashion. For each new redshift a corresponding fit provides the new LC parameters (stretch, date of maximum and flux scales) and their respective errors (and covariance matrix). These values and errors, in turn, are used to generate a new set of random LC values (for SNe with and without photo-$z$) that determine the model SED for each epoch. The $K$-corrections are recalculated for the new set of redshift and LC parameters. For each of these realizations we fit the rise-time data of all SNe to the model and get the best fiducial, stretch-corrected rise-time $t_r$ and slope $a$, respectively. The final values are drawn from the median of all realizations combined. The statistical uncertainties in the parameters are given by the 1$\sigma$ uncertainties in the distribution (see Figure~\ref{mcrisehist}). The distribution is broader for low-$s$ because of the smaller sample size,and because the low-$s$ sample includes more SNe with photometric redshifts, and hence larger errors in $z$ (see G11 for details).

Final rise-times for the SNLS, SDSS-II and nearby samples are shown in Table~\ref{table_fits}. The different samples agree well with each other providing no evidence for evolution. This is confirmed by the fiducial rise-time of the SNLS SNe~Ia at $z>0.7$. Rise-times for different SNLS stretch populations after the primary LC-width, or stretch, correction are also consistent. However, the data are also in agreement at a $\sim1\sigma$ level with a secondary slight trend for longer \emph{fiducial} rise-times in fast decliners. High-stretch SNe ($s>1.0$) have the shortest stretch-corrected rise-times, $\sim1\sigma$ shorter than mid-stretch SNe ($0.8<s\leq1.0$), and low-stretch ($s\leq0.8$) also seem to follow this trend. In the next sections, we investigate systematic effects that could modify these results.

\subsection{Comparison of single- and two-stretch fits}\label{2s}

\begin{figure}[htbp]
     \centering
         \includegraphics[width=1.0\linewidth]{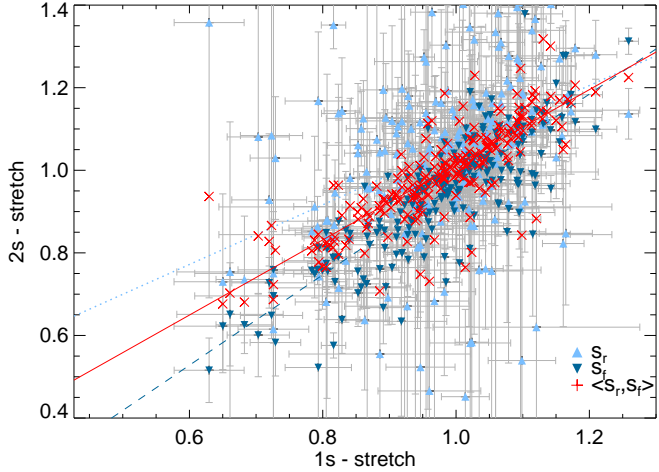}
         \caption{Comparison of rise- (light blue triangles) and fall-stretch (dark blue inverted triangles) for SNLS SNe~Ia calculated with the two-stretch technique as a function of the single-stretch. The crosses are the mean of the rise- and fall-stretches. The lines are linear fits: solid (red) for the mean of rise- and fall-stretches (slope: $0.91\pm0.08$), dotted (light blue) for rise-stretch (slope: $0.72\pm0.07$) and dashed (dark blue) for the fall-stretch (slope: $1.09\pm0.04$).}
         \label{twost-study}
       \end{figure}

Here, we investigate differences arising from the parameterization of the LC in two pieces (rising and falling) as in Equation~\ref{2s-eq}, and as done in H10. We process our data by fitting it with a 2-$s$ model, and we find that the rise-stretch and fall-stretch are consistent within the errors, although the average rise-stretch is slightly larger than the fall-stretch: $\left<s_r-s_f\right>=0.09$ (RMS of 0.24). The mean rise-stretch is also larger than the stretch from the 1-$s$ model, $\left<s_r-s\right>=0.06$ (RMS of 0.17) (see Figure~\ref{twost-study}). This can be explained as the day of maximum gets pushed to later times by more than half a day, $\left<t_0(s_r,s_f)-t_0(s)\right>=0.65$ (RMS of 1.77). 

\begin{figure}[htbp]
     \centering
         \includegraphics[width=0.75\linewidth]{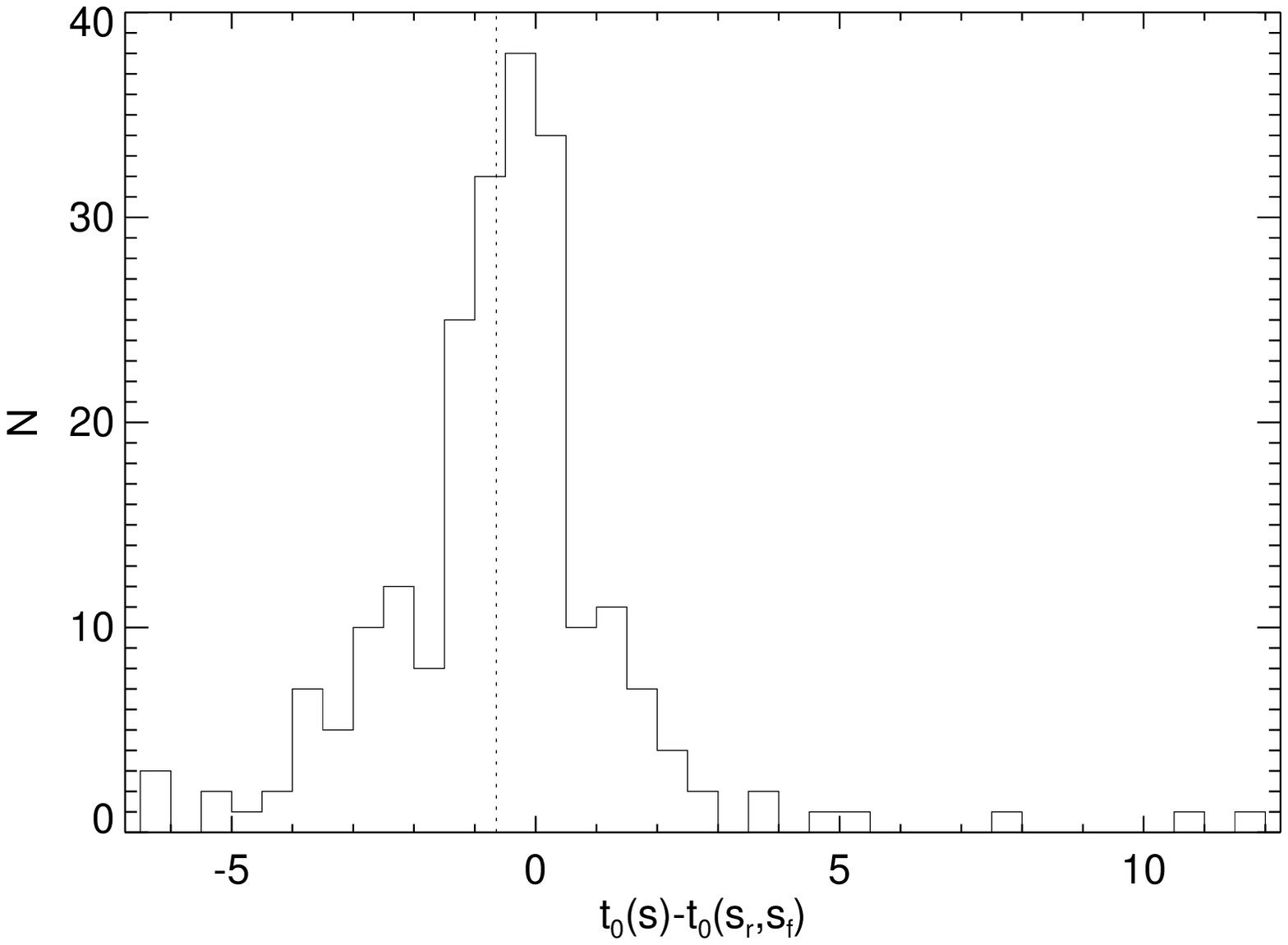}
         \includegraphics[width=0.75\linewidth]{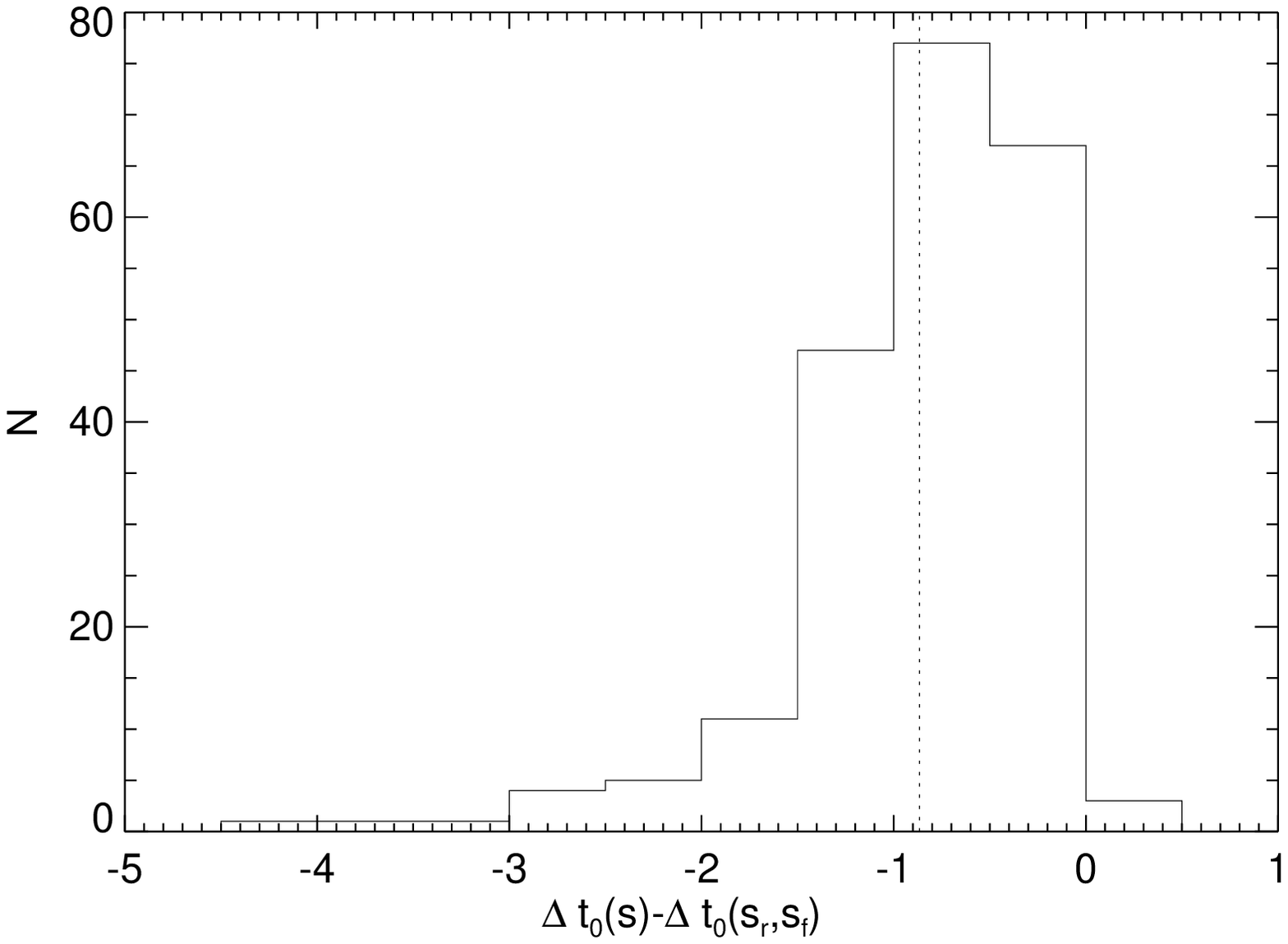}
         \caption{Difference of peak date (upper) and peak date error (lower) calculated with the 1-$s$ and the 2-$s$ techniques for the SNLS sample}
         \label{tmax}
       \end{figure}

The mean error in the peak date is much larger for the 2-$s$ fits, $\left<\Delta t_0(s_r,s_f)-\Delta t_0(s)\right>=0.87$ (RMS of 0.53) (see Figure~\ref{tmax}). An incorrect estimate of the date of maximum has a direct effect on the rise-time calculation. For our MC technique, in which the fit parameters are varied according to their errors, large uncertainties in the peak epoch will affect the final rise-time even more. Our initial cut of $\Delta t_{0}<0.7$ days rejects more than 80\% of the 2-$s$ fits of the SNLS sample (40\% for the 1-$s$ fits), including all low-$s$ objects and all $z>0.7$ SNe~Ia. We therefore relax the cut to $\Delta t_0<1.2$ days for the 2-$s$ fits.

We do not change our low-$s$ definition for the 2-$s$ fits to $s_r\leq0.8$ as this would reflect only the early part of the LC. If we were to use the fall-stretch instead, some low-$s$ candidates would have even lower stretch values. Instead we note that most candidates have a mean 2-$s$ stretch, i.~e. $(s_r+s_f)/2$, closer to the single-stretch: $(s_r+s_f)/2-s=0.01$ with 0.05 RMS, and use this to separate the different stretch samples. This definition keeps most of the low-$s$ candidates and is the best overall description of the lightcurve.

With the new stretch values we derived for the 2-$s$ fits, we repeat the procedure to estimate the fiducial rise-time, using the rise-stretch instead of the single stretch values for the $K$-correction and normalization of all pre-maximum epochs. We obtain shorter stretch-corrected rise-times than the 1-$s$ approach, about 1-2\% for the different SNLS samples (5\% for SNe~Ia at $z>0.7$), but still within the uncertainties. The decrease in fiducial rise-time with the 2-$s$ respect to the 1-$s$ fit comes partially from the compensation for the higher rise-stretches measured. 

The 2-$s$ technique introduces new parameters in the fit, and we want to assess if an improvement is simply due to the increased complexity of the model, or if there is a real advantage in using it. To compare the two models, we use the Bayesian Information Criterion, $\mathrm{BIC}=\chi^2+f\ln(N)$, a tool that gives weight to the number of parameters of the fit, $f$, as well as the total data used, $N$ \citep{Schwarz78}. By summing all the individual core LC fits ($-10<\tau<35$ days) of all SNe, we find that the 1-$s$ fit gives a smaller BICvalue: $\mathrm{BIC}=17405$, compared to the 2-$s$ value: $\mathrm{BIC}=19721$ (the average $\chi^2$ per degree of freedom for the individual 2-$s$ fits is slightly better, $\left<\chi^2_{\nu}\right>=1.44$, than for the 1-$s$ technique: $\left<\chi^2\right>=1.48$), i.~e. the 2-$s$ technique did not lead to a better fit once the increase in parameters is taken into account. If we only consider the rise portion of the lightcurve (for which we use the same quadratic fit for both techniques), we obtain a lower chi-square per degree of freedom for the 1-$s$ approach, $\chi^2_{\nu}=1.60$, than for the 2-$s$ fit: $\chi^2_{\nu}=1.64$. The SNLS systematic uncertainties from different LC fits, 2-$s$ and a newer SiFTO version different from G11, are shown in Table ~\ref{table_syst}. 

Similar results are found for the SDSS and low-$z$ samples. The core LC fits are better for the 1-$s$ fits whereas the early rise-time fit is better for the 2-$s$ approach, according to the BIC criterion. The average rise-stretch is larger than the fall-stretch (by $0.09$ for the SDSS and $0.11$ for the low-$z$ sample) partly because of a mean shift in the peak date (of $0.45$ and $0.40$ days for the SDSS and low-$z$, respectively). The mean errors in peak time are somewhat lower ($0.69$ for the SDSS and $0.21$ for the nearby sample). The final rise-times obtained are comparable for both techniques, although the low-$z$ 2-$s$ result is $\sim1.5\sigma$ lower than its 1-$s$ counterpart. 

In the present study, we do not find a significant change in the rise-times of different samples for the two parameterizations considered: 1-$s$ and 2-$s$, except for the nearby set for which a $\sim1.5\sigma$ difference is found. Our 2-$s$ technique that is based on an adapted two-piece SiFTO fitter provides more uncertain peak date estimates. Requiring accurate maximum epochs affects the number of SNe we can use in the fit for the 2-$s$ technique (see section~\ref{data}). We find that using only SNe with $\Delta t_0<1.2$ days is good enough for the fit but the number of objects used is less than for the 1-$s$ technique, even though we use a tighter criterium for that fitter of $\Delta t_0<0.7$ days. 

Although the statistical analysis of the fiducial rise-time remains unchanged with the parameterization used, Figure~\ref{twost-study} shows that individual rise- and fall-stretches in the 2-$s$ technique do not vary equally for different SNe. The linear fits show that lower-$s$ SNe~Ia tend to have larger rise-stretches and shorter fall-stretches for their respective 1-$s$ (a higher difference between the two) as compared to higher-$s$ SNe~Ia. The slopes of rise- and fall-stretch are different by more than 3$\sigma$.

\begin{table*}[htbp]
 \centering
\small
\caption{Stretch-corrected rise-time fit results for the single- and two-stretch lightcurve fits (with $\tau_t=-10$ days)}
 \begin{threeparttable}
 \renewcommand{\tabcolsep}{3.5mm} 
   \renewcommand{\arraystretch}{1.15}
  \begin{tabular}{|cc|c|c|c|c|}
    \hline
    \hline
    \multicolumn{2}{|c|}{Sample} & $N_{SN}$(1-$s$) & $t_r$(1-$s$) (days)\tnote{*} & $N_{SN}$(2-$s$) & $t_r$(2-$s$) (days)\tnote{*} \\
    \hline \hline
    SNLS $z<0.7$ & & 134 & $17.02^{+0.19}_{-0.26}$ & 87 & $16.92^{+0.38}_{-0.49}$ \\
    & $s\leq0.8$ & 12 & $18.03^{+0.81}_{-1.37}$ & 8 & $17.89^{+1.31}_{-2.33}$ \\
    & $0.8<s\leq1.0$ & 55 & $17.44^{+0.45}_{-0.43}$ & 31 & $17.56^{+0.63}_{-0.77}$ \\
    & $s>1.0$  & 67 & $16.87^{+0.15}_{-0.25}$ & 48 & $16.62^{+0.44}_{-0.58}$ \\
    \hline
    SNLS $z>0.7$ & & 102 & $17.11^{+0.24}_{-0.36}$ & 15 & $16.11^{+0.68}_{-0.77}$\\
 \hline
    Nearby & & 16 & $17.39^{+0.16}_{-0.24}$ & 15 & $16.58^{+0.21}_{-0.27}$ \\
    \hline
    SDSS-II& & 64 &  $17.44^{+0.33}_{-0.53}$ & 36 & $17.03^{+0.26}_{-0.40}$ \\ 
    & $0.8<s\leq1.0$ & 23 & $17.78^{+0.34}_{-0.40}$ & 12 & $17.51^{+0.31}_{-0.68}$ \\
    & $s>1.0$  & 41 & $17.37^{+0.44}_{-0.44}$ & 24 & $16.91^{+0.24}_{-0.48}$ \\
    \hline
   \end{tabular}
\begin{tablenotes}
\item [*]  Statistical MC errors only
\end{tablenotes}
\end{threeparttable}

\label{table_fits}
\end{table*} 

\begin{table*}[htbp]
\centering
 \caption{Stretch-corrected rise-time systematic uncertainties for the SNLS stretch samples: low-$s$ ($s\leq0.8$), mid-$s$ ($0.8<s\leq1.0$) and high-$s$ ($s>0.8$) and differences between them}
 \begin{threeparttable} 
   \renewcommand{\arraystretch}{1.5}
  \begin{tabular}{|c|cccc|cc|}
    \hline
    \hline

{\bf Systematic} & \multicolumn{4}{c|}{\bf Rise-time $t_r$ (days)} & \multicolumn{2}{c|}{\bf Difference $\Delta t_r$ (days)} \\ 
  & all & low-$s$ & mid-$s$ & high-$s$ & low-$s$ - mid$s$ & mid-$s$ - high-$s$ \\
  &     &($s\leq0.8$)& ($0.8<s\leq1.0$) & ($s>1.0$) & & \\
    \hline

    {\bf STANDARD}  & $17.02^{+0.19}_{-0.26}$ & $18.03^{+0.81}_{-1.37}$ & $17.44^{+0.45}_{-0.43}$ & $16.87^{+0.15}_{-0.25}$ & $0.59^{+0.93}_{-1.44}$ & $0.57^{+0.47}_{-0.50}$ \\ 
    \hline
    {\bf  LC-FIT} & & & & & & \\
    1-$s$ new SiFTO      & 17.14 & 18.46 & 17.99 & 16.57 & 0.47 & 1.42 \\ 
    2-$s$                & 16.92 & 17.89 & 17.56 & 16.62 & 0.97 & 0.94 \\ 
    \hline

    \hline
    {\bf TEMPLATE} & & & & & & \\
    NEW (conley09f)& 16.94 & 18.33 & 17.70 & 16.59 & 0.63 & 1.11 \\ 
    C06 (nugent05)\tnote{*} & 18.77 & 20.13 & 19.68 & 18.21 & 0.45 & 1.47 \\ 
    SALT2          & 18.16 & 19.13 & 18.77 & 17.57 & 0.36 & 1.20 \\
    \hline

      {\bf  TRANSITION EPOCH} & & & & & & \\
   $\tau_t=-8$d   & 17.22 & 18.68 & 17.56 & 16.95 & 1.12 & 0.61 \\
  \hline

   {\bf  PHOTOMETRY} & & & & & & \\
    Fr phot, $b=0$    & 17.21 & 17.96 & 17.40 & 16.96 & 1.02 & 0.44 \\ 
    Ca phot, free $b$ & 17.18 & 18.07 & 17.50 & 16.86 & 1.03 & 0.64 \\ 
    \hline

  {\bf  POWER LAW} & & & & & & \\
    $n=1.9$ & 16.83 & 17.87 & 17.26 & 16.68 & 0.61 & 0.58 \\ 
  \hline \hline 
 $\boldsymbol{\sigma_{\mathrm{syst}}^{\mathrm{tot}}}$ & $\boldsymbol{^{+1.78}_{-0.23}}$ & $\boldsymbol{^{+2.24}_{-0.22}}$ & $\boldsymbol{^{+2.31}_{-0.18}}$ & $\boldsymbol{^{+1.35}_{-0.45}}$ & $\boldsymbol{^{+0.69}_{-0.27}}$ & $\boldsymbol{^{+1.24}_{-0.13}}$\\

\hline            
  \end{tabular}
\begin{tablenotes}
\item [*] This template is based on an old limited SN sample and is not included in the final systematic error budget.
\end{tablenotes}
\end{threeparttable}
\label{table_syst}
\end{table*}

\subsection{Comparison of different templates}\label{template}


The template used in C06 descended from \citet{Nugent02} and the early portion was based on a limited number of SNe~Ia. Using this template to calculate the rise-time, we obtain 19.33 days, in agreement with the result obtained in C06 of 19.10 days.

\begin{figure}[htbp]
        \centering
        \includegraphics[width=0.49\linewidth]{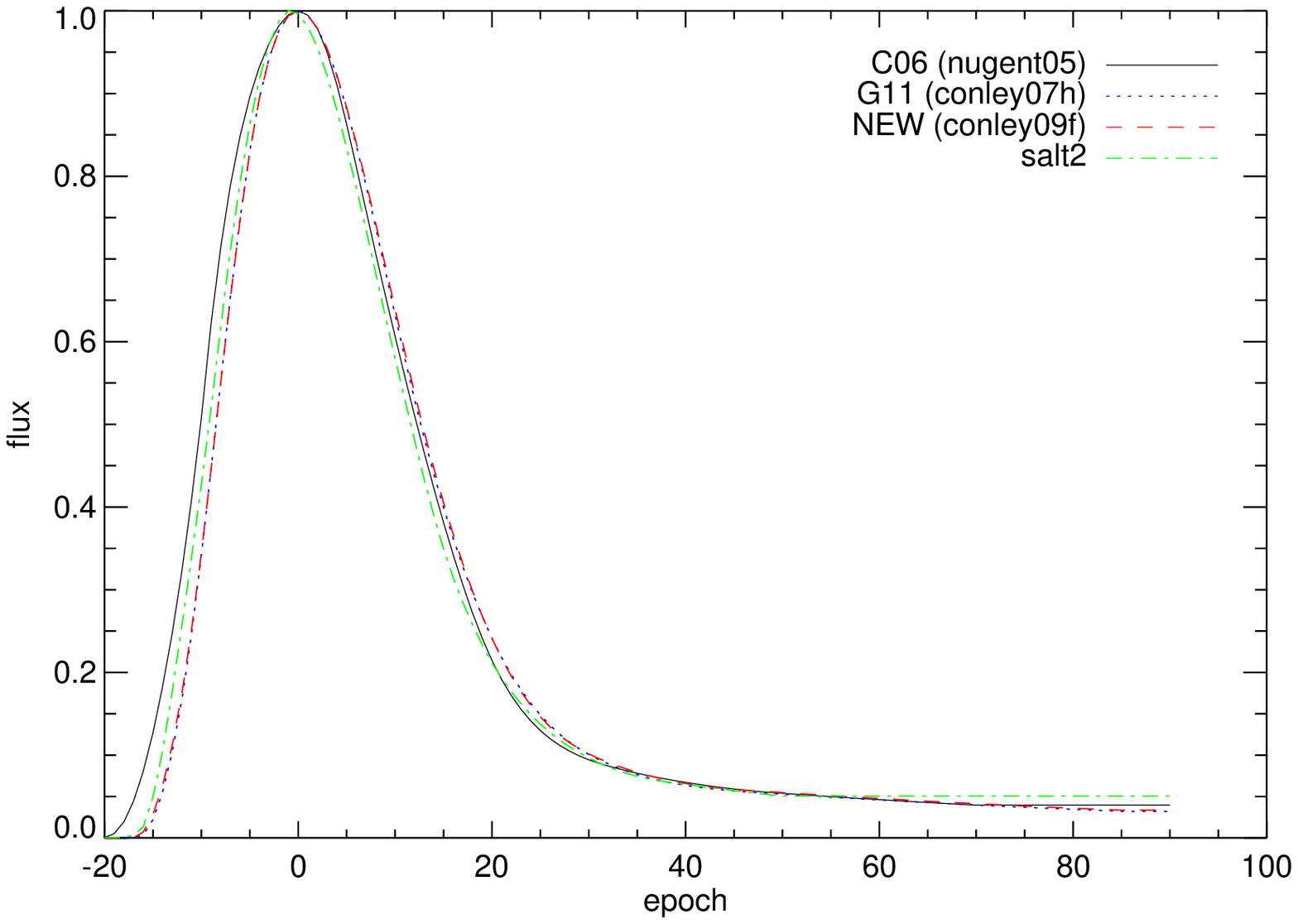}
        \includegraphics[width=0.49\linewidth]{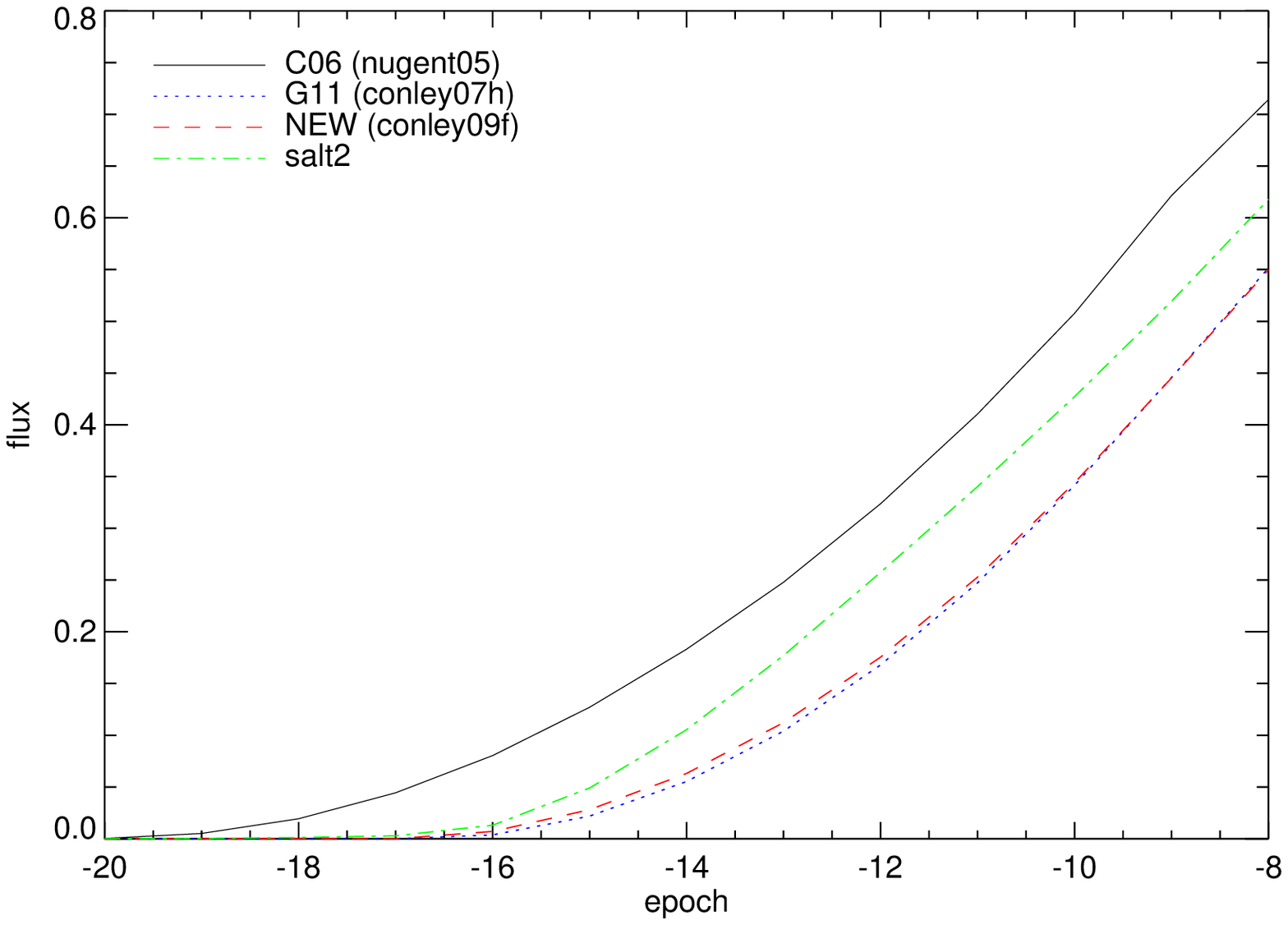}
        \caption[Different SN~Ia light-curve templates]{Different SN~Ia light-curve templates (inset of early portion on the right): C06 (solid black), G11 (dotted blue), NEW (dashed red) and SALT2 (dot-dashed green).}
        \label{templating}
      \end{figure}

\begin{figure}[htbp]
        \centering
        \includegraphics[width=0.9\linewidth]{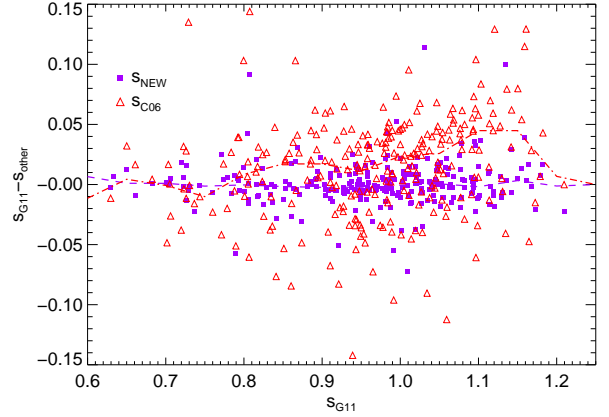}
        \caption{Stretch comparison between lightcurve fits of the G11 template and other templates: C06 (red triangles), and a more recent one NEW (purple squares). The red dot-dashed and purple dashed lines are the averaged median in 0.05 stretch bins.}
        \label{temp-comp}
      \end{figure}

The fiducial stretch-corrected SN~Ia rise-time is dependent on the SED template used (see Figure~\ref{templating}), in particular on the rather arbitrary definition of a $s=1$ SN. This is especially important when comparing rise-times from different sources, where various templates can lead to different rise-times. A $s=1$ SN is defined here as the mean stretch of the training sample for the template we use, which is the same as in G11. If we correct rise-time of the C06 template for the different $s=1$ definitions, by comparing the fits directly (see Figure \ref{temp-comp}), i.~e. a $s=1$ SN for the G11 template corresponds to a $s=1.03$ for the C06 template, we then obtain 18.77 days. This averaged correction partially accounts for the early wide behavior of the C06 template (see Figure~\ref{overLC-early}). The simple change of $s=1$ definition is not sufficient to fully compensate the effect of an incorrect early template, so that a longer stretch-corrected rise-time is still obtained. We get similar long rise-times with the C06 template applied to the 2-$s$ technique, indicating that this parameterization is not independent of the template used.

Newer templates trained with the well-sampled data of the SNLS, like the one used in G11, or a more recent version trained with a larger sample (used for SiFTO in \citealt{Guy10}), give consistent rise-times after correction for the different $s=1$ definition. Table~\ref{table_syst} shows the rise-times for different templates. 

\subsection{Comparison with SALT2}\label{salt2}
We investigate here the use of a different LC fitter, SALT2 \citep{Guy07}. This fitter also makes use of different templates from the ones presented in the previous section, so that we are actually probing two systematics simultaneously: template and LC fitter.

\begin{figure}[htbp]
        \centering
        \includegraphics[width=0.7\linewidth]{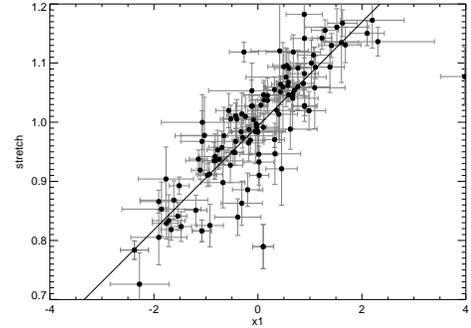}
        \caption{Comparison of SiFTO stretch and SALT2 $X_1$ for the SNLS SN~Ia sample. The solid line represents the fit: $st=a\times X_1+b$ with $a=0.086\pm0.026$ and $b=0.997\pm0.022$.}
        \label{salt2comp}
      \end{figure}

In order to do this, we fit all SNe~Ia with SALT2 in the range $\tau>-10$ days. A direct comparison of their SiFTO stretch and $X_1$ parameter from SALT2 is shown in Figure~\ref{salt2comp} through a linear fit. We calculate restframe, stretch-corrected epochs using the redshift and a corresponding stretch (for each X1 parameter): $\tau=(t-t_0)/s(X_1) (1+z)$. Every data point matching restframe $B$-band is $K$-corrrected by taking the SED for each restframe epoch based on Eq.~1 of \citet{Guy07} with $X_0$, $X_1$ and the color obtained from the SALT2 fit, and integrating it in the appropriate filter at the proper redshift. We then fit the aggregate LC points of the early rise ($\tau<-10$ days) to the quadratic rise model.

Similarly, we take into account errors with a MC that shifts the fit parameters ($z_{\mathrm{phot}}$ if necessary, and $\tau_{\mathrm{max}}$, $X_0$, $X_1$ and color) according to their covariance matrix, and re-calculate a stretch from this, as well as a new K-correction for each point of each simulation. The median of all iteriations, the final fiducial rise-time we obtain for the normal-$s$ population, is $18.16\pm0.44$ days. This is larger by $\sim1.1$ days than the standard SiFTO 1-$s$  and 2-$s$ rise-times. Although the SALT2 template is different than the templates we use (see Figure~\ref{templating}), it is not as extreme as the C06 template. Clearly, the use of the LC fitter also affects the measured rise-time and is an important systematic. The differences between SN~Ia populations are still consistent as shown in Table~\ref{table_syst}.

\subsection{Transition Epoch}\label{transition}
\begin{figure}[htbp]
        \centering
                \includegraphics[width=0.9\linewidth]{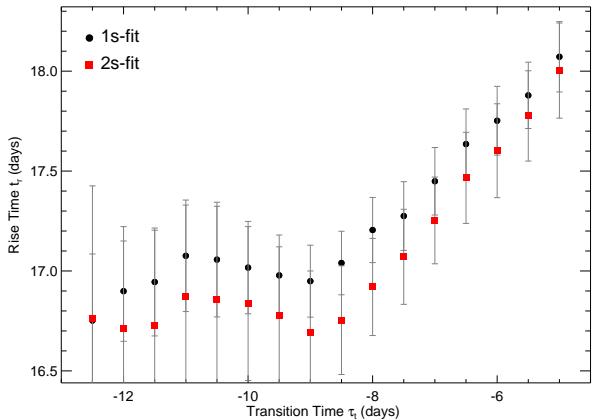}
                \caption{Stretch-corrected rise-time as a function of the transition epoch, the maximum date allowed in the single- (black circles) and two-stretch (red squares) rise-time fits. The rise-time errors are the MC errors calculated as in \S~\ref{rise-calc}. The larger points at $\tau_t=-10$ days are the quoted fiducial rise-times of this study.}
        \label{trans}
\end{figure}

The transition epoch is the minimum epoch used in the core LC fit and the maximum epoch allowed in the early rise-time fit. Its value can affect the quality of the fit but also the resulting rise-time (and even the quadratic behavior of the rise). A higher value, like $\tau_t=-6$ days, increases the amount of data for the early-time fit, but the rise may no longer follow a simple quadratic relation. Figure~\ref{trans} shows that rise-time fits with different transition times and a fixed power fit (quadratic $n=2$) for the SNLS do not change much up to a maximum value of $\tau_t\simeq-8$ days, where the rise-time starts to diverge. We use here the standard $\tau_t=-10$ days used in other studies but add a systematic uncertainty to the rise-time based on the differences at $\tau_t=-8$ days (see Table~\ref{table_syst}).   

\subsection{Zero scale $b$}\label{beta}
Wrong photometric subtraction affects fluxes that are close to zero and strongly influences the final stretch-corrected rise-time calculation. To study this systematic error, we add a nuisance parameter, $b$, to the quadratic form (Eq.~\ref{quadrise}) that will absorb any general photometric error.  For a good photometry, we should obtain $b$ values close to zero. In our study, we get $b=(-0.48^{+0.12}_{-0.09})\times10^{-3}$, an off-zero value reflecting the large dispersion of the fluxes near zero (median of $3.1\times10^{-4}$ with 0.03 RMS) but small enough to represent a negligible effect on the rise-time. Consistent rise-times are obtained with free/fixed $b$ values for the photometry processed via the French and Canadian reconstructions, as seen in Table ~\ref{table_syst}.

The importance of this parameter can be observed in Figure~\ref{add}, in which different $b$ values in the range $-0.05<b<0.05$, or a 5\% maximum shift of the normalized flux at peak, result in rise-times shifted by several days. We emphasize therefore the importance of proper photometry, well-sampled to dates even before explosion, to correctly calculate $b$ and not affect the rise-time.

\subsection{Different rise power laws}\label{power}
Other parameters that could influence the fiducial rise-time behavior are $a$ and the rise power $n$ in Eq.~\ref{quadrise}. We find $a=(7.40^{+0.58}_{-0.62})\times10^{-3}$ for the normal-$s$ SNLS population (arbitrary flux units) and an assumed $n=2$ quadratic form. If we relax the quadratic rise assumption and allow the power of the rise-time model to be free (free $a=(11.62\pm0.85)\times10^{-3}$ and fixed $b=0$), we obtain $n=1.92^{+0.31}_{-0.37}$(stat). Rise-times fitted with this new power law are lower for all stretch populations but consistent with their quadratic equivalents.

\begin{figure}
     \centering
     \subfigure{\label{add}
       \includegraphics[width=.45\textwidth]{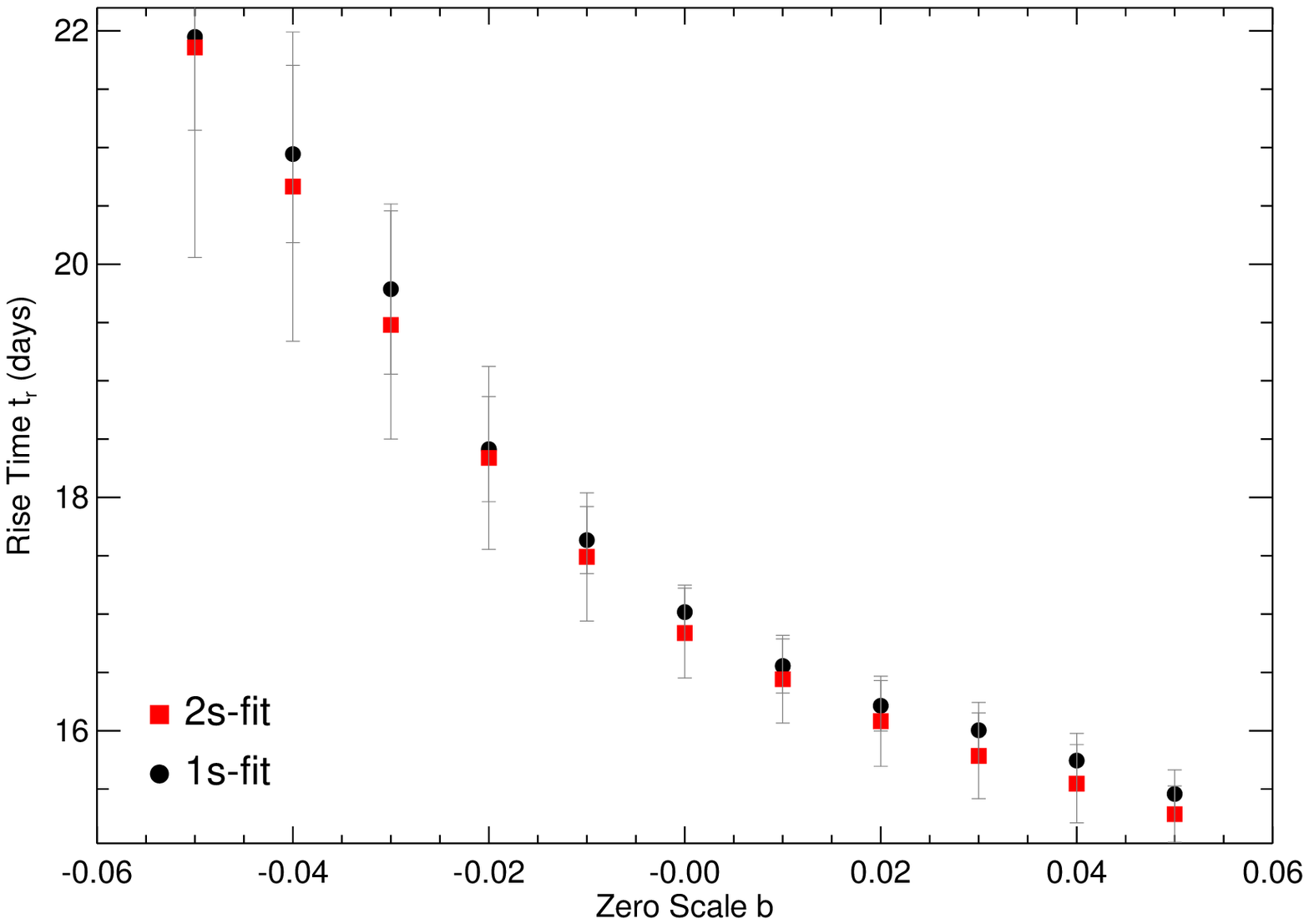}}
     \hspace{.3in}
     \subfigure{\label{pow}
       \includegraphics[width=0.45\textwidth]{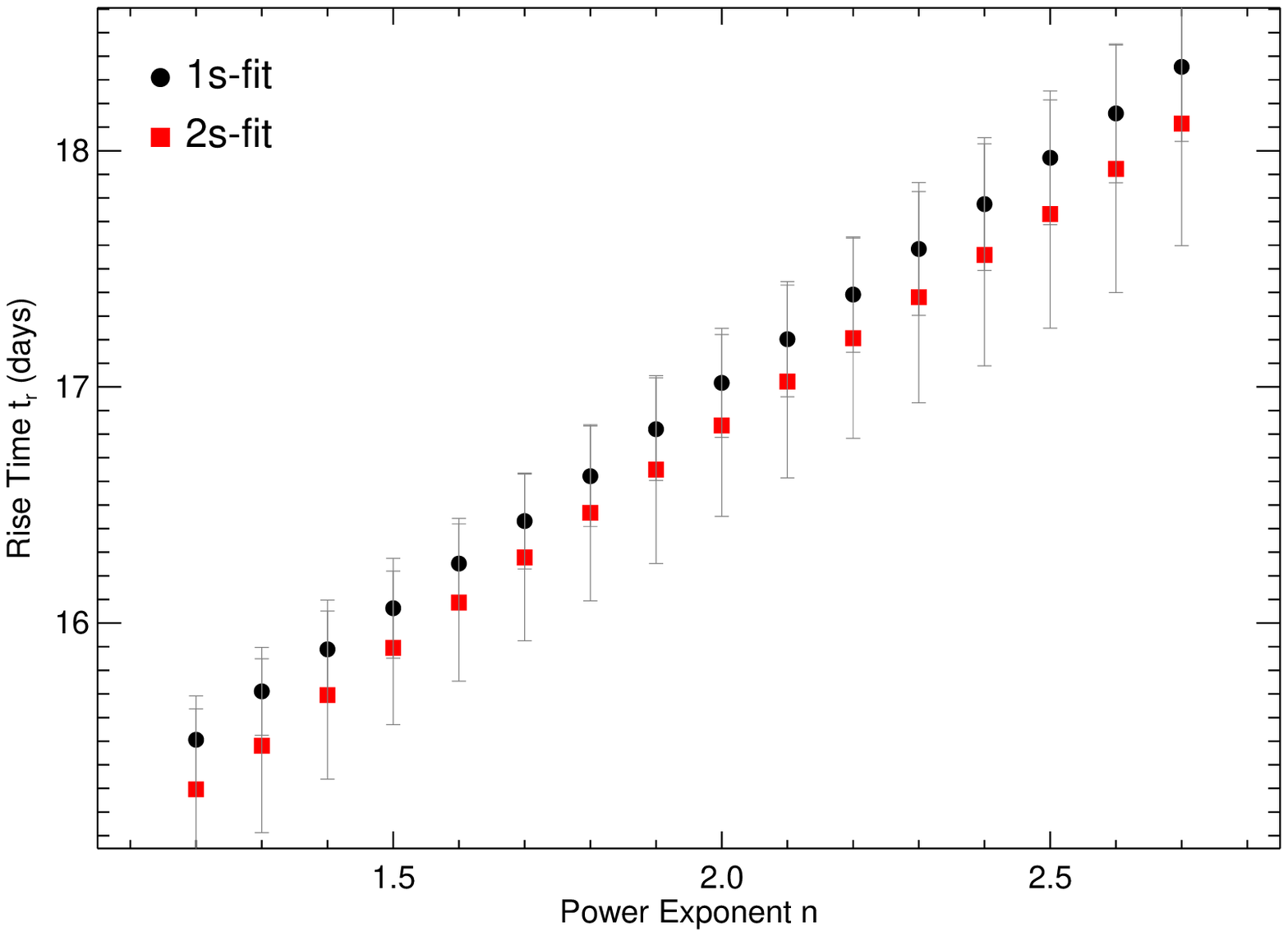}}
     \caption{Rise-time $t_r$ as a function of zero scale $b$ (\emph{upper}) and power exponent $n$ (\emph{lower}) of Equation~\ref{quadrise}, as obtained in the fits for 1-$s$ (black circles) and 2-$s$ (red squares) techniques (at $\tau_t=-8$ days). Errors indicate MC uncertainties (\S~\ref{rise-calc}). Empty points are the best fits with $b$ and $n$ free, respectively ($a$ was kept free for all fits).}
\end{figure}

Figure~\ref{pow} shows the relation between the assumed power-law, $n$, and the measured stretch-corrected rise-time, $t_r$ for the normal SN~Ia sample. A higher $n$ makes the LC rise faster. In order to compensate for this, the duration of the rise to maximum of the fit becomes longer. This effect has a high degree of degeneracy and correlation between the two fit parameters, $t_r$ and $n$. A very similar trend is observed for the different stretch samples. Furthermore, if the quadratic form is relaxed for all populations, we obtain increasing power laws and rise time for higher-$s$ SNe (Table ~\ref{table:powexp}). This variation shows again the clear difference in the rise-time region for the sub-samples of SNe~Ia. 

\begin{table}[htbp]
 \centering
\caption{Stretch-corrected rise-time and power-exponents fit results for the different stretch samples in the SNLS}
\begin{threeparttable}
\renewcommand{\tabcolsep}{3.5mm} 
   \renewcommand{\arraystretch}{1.5}
 \begin{tabular}{|c|c|c|}
    \hline
    \hline
    Sample & $t_r$(days)\tnote{*} & $n$\tnote{*} \\
    \hline
    SNLS & $16.85^{+0.54}_{-0.81}$ & $1.92^{+0.31}_{-0.37}$ \\
    High-$s$ & $17.89^{+0.77}_{-0.98}$ & $2.66^{+0.44}_{-0.59}$ \\
    Mid-$s$  & $16.49^{+0.56}_{-0.88}$ & $1.48^{+0.34}_{-0.37}$ \\
    Low-$s$  & $16.14^{+1.00}_{-1.99}$ & $0.78^{+0.71}_{-0.71}$ \\ 
 \hline
  \end{tabular}
\begin{tablenotes}
\item [*]  Statistical MC errors
\end{tablenotes}
\end{threeparttable}
\label{table:powexp}
\end{table}            

Although the systematic uncertainty from this variation is also highly correlated with the transition time, we still include it in the total systematic error budget. 




\section{Results and Discussion}\label{discussion}


The final fiducial rise-time, $t_r$, for all SNe~Ia in the SNLS calculated via the single-stretch method is $t_r=17.02^{+0.19}_{-0.26}$(statistical)$^{+0.30}_{-0.23}$(systematic). The statistical uncertainty represents the 1$\sigma$ distribution of the MC simulations and the systematic uncertainty comes from the different effects considered in the previous section (added in quadrature) and shown in Table~\ref{table_syst}. This agrees with the value found via the two-stretch technique. 

We find that the fiducial stretch-corrected rise-times are consistent for fast and slow decliners, independently of different fit characteristics such as the template or lightcurve fitter. Mid-stretch SNe~Ia ($0.8<s\leq1.0$) have a fiducial rise-time $\Delta t_r=0.57^{+0.47}_{-0.50}$(stat)$^{+1.24}_{-0.13}$(syst) days longer than their high-stretch counterpart ($s>1.0$), barely a $\sim1\sigma$ result. With even lower significance, low-stretch SNe~Ia show a difference with respect to the high-stretch population of $\Delta t_r=1.16^{+0.82}_{-1.39}$(stat)$^{+1.42}_{-0.19}$(syst) days. 

Although we do not measure individual rise-times, an estimate based on the stretch of each object can be obtained from the fiducial rise-time. For the single-stretch case we do $\tau_{r,i}=s_i\times t_r(s)$, where the subscript $i$ denotes each individual SN and $t_r(s)$ is the fiducial rise-time. Individual rise-times for the low-$s$ population lie in the range of 10-14 days. For the normal-$s$ population, they lie in the range of 14-21 days.

\subsection{Comparison to previous studies}\label{previous}
As mentioned in \S~\ref{template}, the different definition of a fiducial $s=1$ lightcurve across different templates makes the comparison of stretch-corrected rise-times from several studies quite difficult. 

Having that in mind, recent studies \citep{Strovink07,Hayden10}, including the present one, agree that the average rise-time of SNe~Ia is shorter than previously thought. With more data, better-sampled at early times, the new templates, as well as LC-fitting techniques, provide evidence for a $\sim$1-2 days shift in the average rise-time of SNe~Ia with respect to previous studies \citep{Riess99b,Conley06}, although the precise value is heavily dependent on the definition of a fiducial $s=1$ LC template. 

Although we use the same technique as C06, the template they used was trained with a handful of SNe~Ia and consequently had a rise-time behavior to maximum quite different from the current templates. Their rise-time is longer than the one we find here, even after correcting for our $s=1$ definition. 

We also analyze the nearby and SDSS-II datasets with our same template and LC fitter. H10 used a much larger SDSS-II sample (361 SNe) than used here (the 146 SDSS-II SNe from \citealt{Holtzman08}). Our SDSS-II fiducial rise-time agrees well with theirs. They argue that the two-stretch technique presents a much better parameterization than a single parameter approach. We do not agree with those findings for our aggregate rise-time fit approach, neither for the SNLS nor the SDSS and low-$z$ data. This could be due to the use of a different fiducial LC template and fitter that differ in the rise-time region from the ones they use (G11 and modified SiFTO vs MLCS2k2, \citealt{Jha07}), in particular the stretch they calculate is not based on multiple bands. The Bayesian Information Criterion for the core LC fits suggest that the inclusion of an additional parameter to model SN~Ia lightcurves does not provide a valuable improvement of the fit. The quadratic fit of the early rise-time data is not strongly affected by the parameterization either. Individual rise-times, on the other hand, could benefit from such an independent parameterization, as shown in Figure~\ref{twost-study}.

Based on our calculations of the fiducial rise-time for different samples across the redshift range $0<z<1$, we do not find evidence for a significant evolution in the early portion of SN~Ia lightcurves. Dividing the SNLS sample into three populations according to their stretch values, we measure the average rise-times for each sub-group and find them to be consistent. Although at low confidence, the data also agree with a trend of faster fiducial rise-times for brighter and slowly declining SNe~Ia \emph{after} the primary stretch correction, for the SNLS and SDSS-II. H10 obtained a similar trend for the individual rise-times of SDSS-II objects.

\begin{figure}[htbp]
        \centering
             \includegraphics[width=1.0\linewidth]{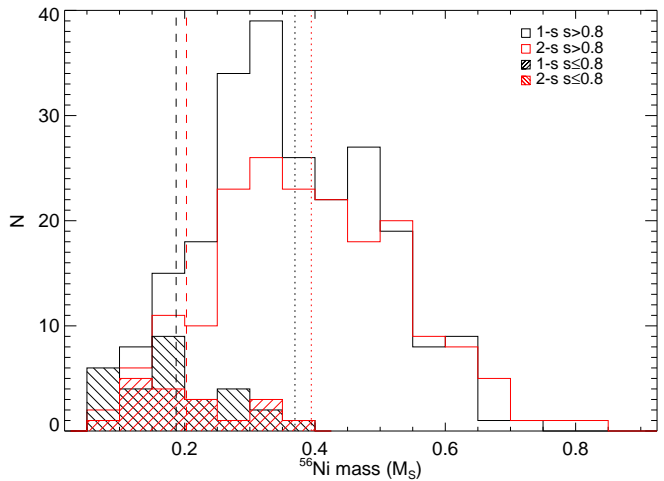}
		\caption{Distribution $^{56}$Ni masses (solar units) for the normal-$s$ (solid histograms) and low-$s$ (filled histograms) SN~Ia SNLS samples based on 1-$s$ (black) and 2-$s$ (red) fits. Vertical lines represent the mean for each case.}
        \label{nimasshist}
\end{figure}

\subsection{$^{56}$Ni masses}\label{nimass}

\begin{figure}[htbp]
        \centering
        \subfigure{
       \includegraphics[width=.45\textwidth]{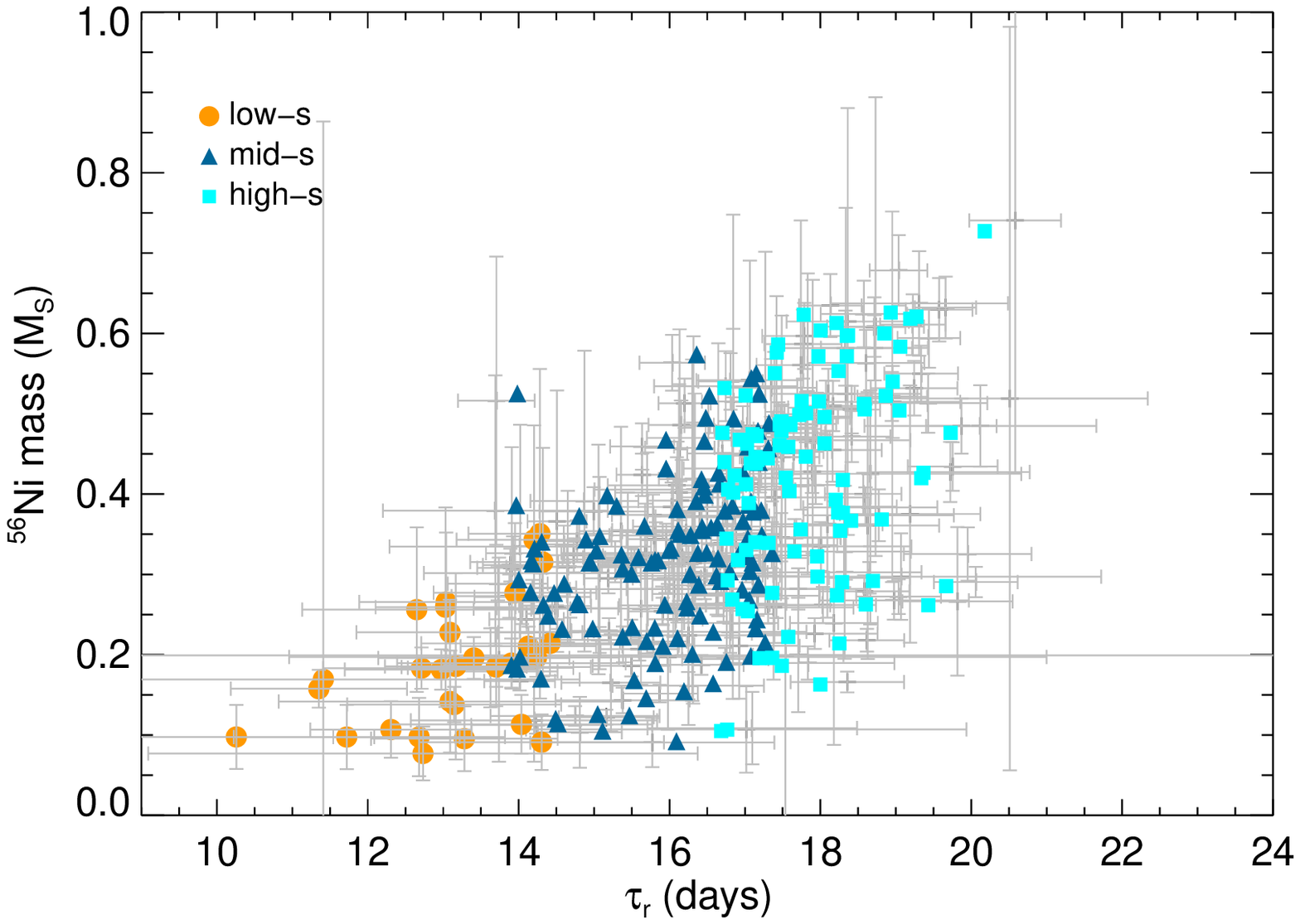}}
     \hspace{.3in}
     \subfigure{
       \includegraphics[width=0.45\textwidth]{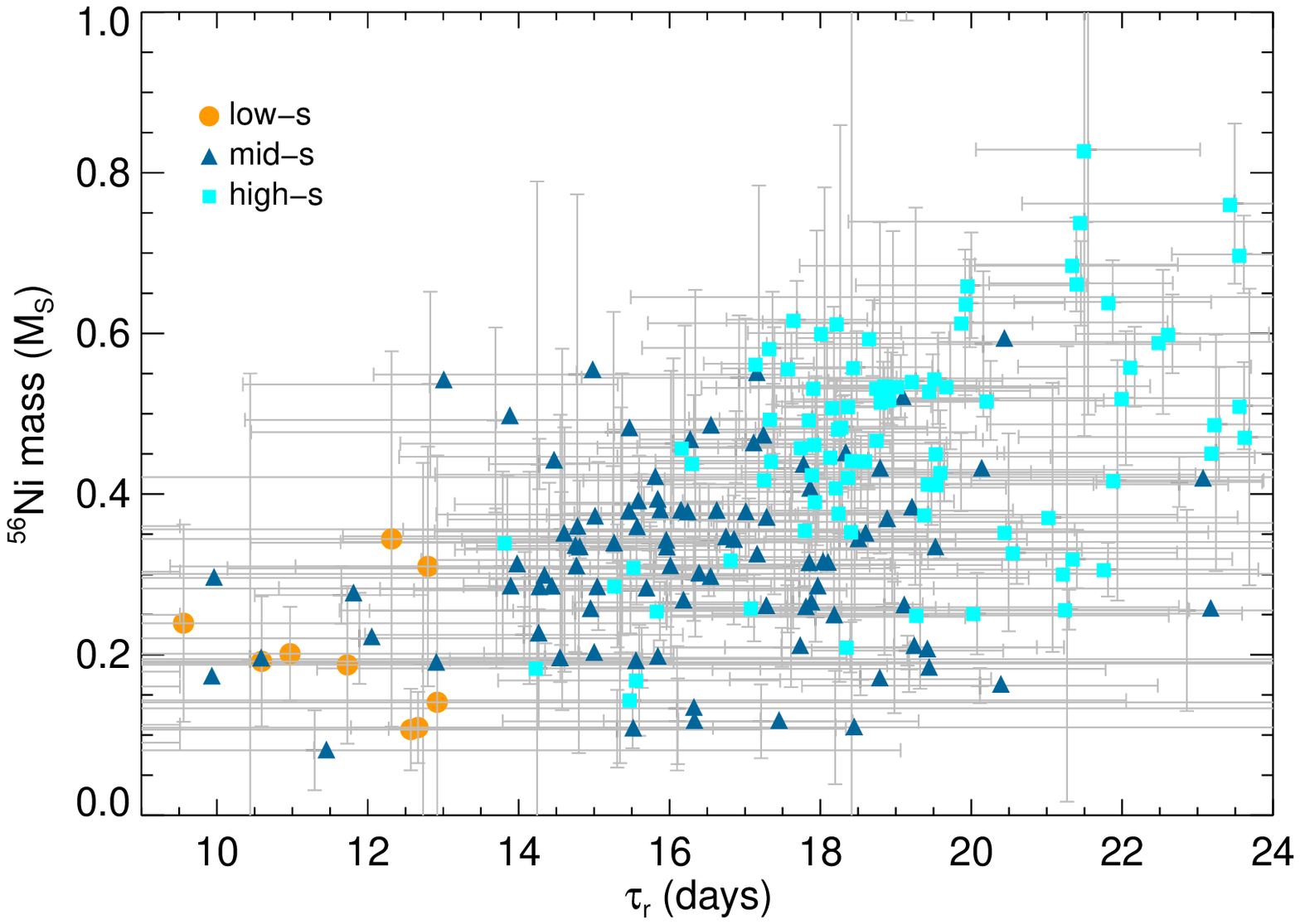}}
       \caption{$^{56}$Ni mass as a function of individual rise-time for the single- (\emph{upper}) and two-stretch (\emph{lower}) techniques. The low-$s$ (orange circles), mid-$s$ (blue triangles) and high-$s$ (light blue squares) SNe~Ia are indicated.}
        \label{nimass-tr}
\end{figure}

In order to estimate the $^{56}$Ni masses, we need to calculate bolometric luminosities. We integrate the model SEDs for each SN across all observed bands. This omits the near-infrared (NIR), which accounts for a substantial fraction of the bolometric luminosity, so we must apply a correction factor. In \citet{Howell09} the IR percentage for normal SNe~Ia is estimated via integration of proper SEDs corrected for each SN in the region outside the observed spectrum, under some assumptions. For low-$s$ SNe~Ia, SEDs are not as fully sampled. We instead use a very broad approximation based on the IR fraction found for SN1999by and SN2004eo \citep{Hoeflich02,Garnavich04,Taubenberger08}: $\sim15-25\%$. The fraction is at least $5\%$ higher than for normal SNe~Ia. We use a standard $20\pm10\%$ value (and $10\pm5\%$ for the normal-$s$) to calculate via Equations~\ref{ni1} and~\ref{ni2} the $^{56}$Ni mass. 

\begin{figure}[htbp]
        \centering
        \includegraphics[width=1.0\linewidth]{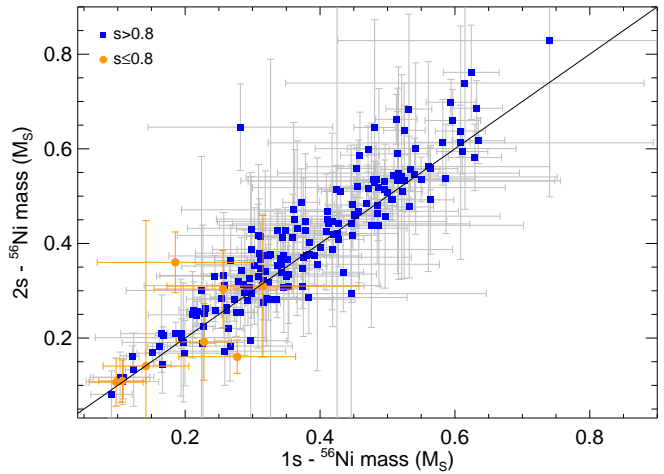}
        \caption{Comparison $^{56}$Ni mass estimates as calculated in \S~\ref{nimass} for the single- and two-stretch techniques. Orange circles are the low-$s$ objects.}
        \label{nimass-comp}
      \end{figure}

In order to account for the errors in the LC parameters, we carry out this computation for every realization of the LC data in the MC of \S~\ref{rise-calc}, and use the median $L_B$ of each SN to calculate the final $^{56}$Ni masses, as shown in Figures~\ref{nimasshist} and ~\ref{nimass-tr}. We also include the error on the conversion factor $\gamma$, which can also slightly vary with $^{56}$Ni mass. The $^{56}$Ni masses lie between 0.05-0.35 $M_{\odot}$ for the low-$s$ and 0.1-0.8 $M_{\odot}$ for the normal-$s$ populations. Our estimates are lower than typical values in the literature because of the lower rise-times. They are similar for both parameterizations of the lightcurve (see Figure~\ref{nimass-comp}).

Figure~\ref{nimass-tr} shows $^{56}$Ni masses as a function of the individual rise-times for all SN~Ia. The relation of Ni mass and rise-time is expected from the use of Arnett's rule. The individual rise-times are obtained from the fiducial rise-time and the individual stretches, so that $^{56}$Ni mass also relates with stretch. However, one can observe that, as rise-time is more independent of stretch for the 2-$s$ technique, there is a more diverse $^{56}$Ni mass range for each stretch population. It may be beneficial to divide the LC into separate regions, rise and fall, to disentangle the rise portion from the primary width-luminosity driver, when studying early-time effects of SNe~Ia.

\subsection{Secondary lightcurve parameters}

The need for secondary parameters to describe SNe~Ia at early times, beyond the traditional core LC parameter like stretch or $\Delta m_{15}$, would have important consequences for the understanding of SNe~Ia. To the precision achieved in this study, these secondary effects in the early rise-time region do not seem necessary to describe SN~Ia LCs. We find a low-significance trend towards longer rise-times in fast decliners \emph{after} stretch correction. If confirmed with more data, this would mean that rise-times do not correlate to second-order with stretch and $^{56}$Ni mass. This would suggest that beyond the radioactive decay of nickel which only sets in later in the LC, some other parameter in the initial explosion in the outer ejecta could play a role. 

Besides a primary parameter responsible for the luminosity-width relation of SNe~Ia, like $^{56}$Ni mass (from a varying density at the transition of subsonic deflagration to supersonic detonation in the delayed-detonation scenario), \citet{Hoeflich10} predict that there could be other variables like the C/O ratio of the WD progenitor that would affect the early part of the lightcurve. Higher C/O ratios result in a smaller binding energy at time of the explosion of the WD. In turn, this leads to higher expansion velocities meaning a faster photosphere recession and a shorter rise to maximum light. Lower main-sequence masses (during He-core and He-shell burning for example) can lead to higher C/O ratios. This would imply that lower-stretch SNe~Ia with longer rise-times come from higher masses. The fact that low-stretch SNe~Ia are primordially found in old environments suggest that their progenitors are long-lived and of low-mass \citep{Gonzalez11}, in contradiction to these findings.

On the other hand, low-stretch SNe having longer rise-times is reminiscent of the behavior predicted by \citet{Fink10} for the double-detonation of sub-Chandrasekhar models. The initial He-shell detonation on the WD surface can lead to a detonation of the CO core of the WD. Varying WD masses (0.8-1.4$M_{\odot}$), and with it their central core and shell densities, give rise to a variety of $^{56}$Ni masses and burning products powering the lightcurve. Particularly they find that brighter events have faster rise-times. This is because brighter explosions produce more core $^{56}$Ni that expand faster with less opacity in the outer shells from iron-group-elements created in the He-shell detonation. Their effect is anti-corrected with stretch, which would indicate a primary correction rather than a secondary one. In this picture, the lightcurve and the rise-time are directly related to the initial mass of the WD. Lower-mass WDs (of $\sim0.8M_{\odot}$) have lower He-shell and CO core densities leading to a distribution of burning products that give rise to slower rise-times for low-stretch SNe~Ia. Lower mass WDs producing fast decliners would agree with the aforementioned observations of older host environments. 


Another interesting feature of the predictions by \citet{Fink10} is that the contribution of other radioactive elements like $^{52}$Fe and  $^{48}$Cr of short decay times can make the rise-time longer. This is particularly true for subluminous SNe~Ia, which in their models produce as much as 20\% of these radioactive nuclei, particularly $^{44}$Ti, which does not contribute to the early lightcurve, but is known to be observed in the spectra as well. These yields are a direct consequence of the initial He-shell detonation.

\section{Conclusions}\label{conclusions}

We have constructed aggregate restframe $B$-band LCs of a sample of spectroscopic and photometrically identified SNe~Ia at $z\leq0.7$ from the SNLS. We obtain fiducial rise-times that are consistent with measurements for the low-$z$ and SDSS-II samples calculated here with the same techniques. We extend our SNLS sample and show that there is no significant evolution in the early portion of SN~Ia lightcurves up to $z=1.0$. 

We find that the use of a two-stretch parameterization for the lightcurve fits of SNe~Ia is equivalent to the single-stretch parameterization for the rise-time study of aggregate lightcurves. We confirm the consistency for the SDSS-II and SNLS samples while the nearby set has a rise-time $\sim1.5\sigma$ lower for the 2-$s$ technique and discrepant with all other calculated rise-times. We insist in the importance of proper lightcurve fitters with parameters that can take into account earlier parts of the lightcurve (stretch vs $\Delta m_{15}$) and that include multi-band information, and of templates trained with sufficient SNe with early time data. We also find indications that the 2-$s$ approach might be useful for detailed studies of individual rise-times, as in \citet{Hayden10,Ganeshalingam11}. 
 

We have compared the early region fits for different stretch population of the SNLS. We find that the stretch-corrected rise-times are consistent within uncertainties for slow and fast decliners. We extend this result to the lowest stretch objects. These findings are independent of systematics such as the maximum epoch allowed in the early time fits, as well as on the two LC parameterizations assumed. The present study averages over samples which can have obvious difficulties like the low statistics of the low-stretch population. At a low significance of $\sim1\sigma$, our results also agree with a trend for longer rise-times in fast declining SNe~Ia. If these indications are confirmed, then SN~Ia explosions differ beyond a single parameter orchestrating the lightcurve of SNe~Ia. Secondary parameters in the early rise region would be necessary to fully model the lightcurves and could provide useful insights into SN~Ia explosions. Such an indication is also found for the SDSS-II, as in H10. 


However, to the precision attained in this study, SNe~Ia remain extremely homogeneous all the way to very early times. One single lightcurve parameter, that possibly translates to a unique physical variable, like the transition density of deflagration to detonation in the delayed detonation model, can account not only for the core lightcurves of SNe~Ia, but largely for the rise-time as well. The similarity of SN~Ia lightcurves (after stretch-correcting) argue for a real common origin, with subluminous SNe~Ia at the extreme end of the population.

The fiducial rise-time we find translates into $^{56}$Ni masses of $0.05-0.90M_{\odot}$ for all SNe~Ia (and $0.05-0.35M_{\odot}$ for low-$s$ SNe Ia). Although our assumed IR contribution to the bolometric luminosity is uncertain, extremely low amounts of radioactive $^{56}$Ni seem enough to power a faint SN~Ia, less than that the amount found by \citet{Filippenko92b,Leibundgut93}. Our SNe with the lowest radioactive nickel yield even rival with peculiar extreme objects such as the ones found by \citet{McClelland10,Kasliwal10,Poznanski10}, and pose a challenge to explosion models.

Such weak explosions seem unlikely to be the outcome of a runaway of a WD near the Chandrasekhar mass. Sub-Chandrasekhar models have been indeed a natural way of explaining subluminous SNe~Ia \citep[e.~g.][]{Livne90,Woosley94}, and possibly all SNe~Ia \citep{Fink10,Sim10,vanKerkwijk10}. Located in passive environments that are characteristic of old populations, low-mass WDs seem a promising progenitor of subluminous SNe~Ia. With recent rejections of a high fraction of red giant donors \citep{Hayden10,Bianco11}, a double degenerate merger leading to the sub-Chandrasekhar explosion of a low-mass WD with very low $^{56}$Ni yield appears increasingly plausible. If SNe~Ia share a common explosion and progenitor mechanism, a varying progenitor WD mass could possibly explain the different SN~Ia radioactive yield and luminosity, and their environment. More and better simulations need to test the validity of this picture. \citet{Pakmor10}, for example, modelled the merger of two $\sim0.9M_{\odot}$ WDs and obtained a subluminous SN~Ia explosion with $\sim0.1M_{\odot}$ of synthetized $^{56}$Ni, in agreement with the bulk of our low-$s$ objects but in disagreement with the expected progenitor masses from the environments. More theoretical development need to complement the abundant current and coming observations of SNe~Ia.

\vspace{20mm}We thank Brian Schmidt for valuable comments. We are grateful to the CFHT Queued Service Observations Team and the entire SNLS collaboration. We acknowledge the support from our funding agencies: NSERC, CIAR, CNRS and CEA. MS acknowledges 
support from the Royal Society.

\bibliographystyle{apj}
\bibliography{astro}

\begin{thebibliography}{86}
\expandafter\ifx\csname natexlab\endcsname\relax\def\natexlab#1{#1}\fi

\bibitem[{{Aldering} {et~al.}(2000){Aldering}, {Knop}, \&
  {Nugent}}]{Aldering00}
{Aldering}, G., {Knop}, R., \& {Nugent}, P. 2000, \aj, 119, 2110

\bibitem[{{Altavilla} {et~al.}(2004)}]{Altavilla04}
{Altavilla}, G. {et~al.} 2004, \mnras, 349, 1344

\bibitem[{{Anupama} {et~al.}(2005){Anupama}, {Sahu}, \& {Jose}}]{Anupama05}
{Anupama}, G.~C., {Sahu}, D.~K., \& {Jose}, J. 2005, \aap, 429, 667

\bibitem[{{Arnett}(1982)}]{Arnett82}
{Arnett}, W.~D. 1982, \apj, 253, 785

\bibitem[{{Arnett} {et~al.}(1985){Arnett}, {Branch}, \& {Wheeler}}]{Arnett85}
{Arnett}, W.~D., {Branch}, D., \& {Wheeler}, J.~C. 1985, \nat, 314, 337

\bibitem[{{Bianco} {et~al.}(2011)}]{Bianco11}
{Bianco}, F.~B. {et~al.} 2011, ArXiv e-prints

\bibitem[{{Boulade} {et~al.}(2003)}]{Boulade03}
{Boulade}, O. {et~al.} 2003, in Society of Photo-Optical Instrumentation
  Engineers (SPIE) Conference Series, Vol. 4841, Society of Photo-Optical
  Instrumentation Engineers (SPIE) Conference Series, ed. {M.~Iye \&
  A.~F.~M.~Moorwood}, 72--81

\bibitem[{{Branch}(1992)}]{Branch92}
{Branch}, D. 1992, \apj, 392, 35

\bibitem[{{Bravo} \& {Garc{\'{\i}}a-Senz}(2008)}]{Bravo08}
{Bravo}, E. \& {Garc{\'{\i}}a-Senz}, D. 2008, \aap, 478, 843

\bibitem[{{Bravo} {et~al.}(2010)}]{Bravo10}
{Bravo}, E. {et~al.} 2010, ArXiv e-prints

\bibitem[{{Colgate} \& {McKee}(1969)}]{Colgate69}
{Colgate}, S.~A. \& {McKee}, C. 1969, \apj, 157, 623

\bibitem[{{Conley} {et~al.}(2006)}]{Conley06}
{Conley}, A. {et~al.} 2006, \aj, 132, 1707

\bibitem[{{Conley} {et~al.}(2008)}]{Conley08}
---. 2008, \apj, 681, 482

\bibitem[{{Conley} {et~al.}(2011)}]{Conley11}
---. 2011, \apjs, 192, 1

\bibitem[{{Contardo} {et~al.}(2000){Contardo}, {Leibundgut}, \&
  {Vacca}}]{Contardo00}
{Contardo}, G., {Leibundgut}, B., \& {Vacca}, W.~D. 2000, \aap, 359, 876

\bibitem[{{Dom{\'{\i}}nguez} {et~al.}(2001){Dom{\'{\i}}nguez}, {H{\"o}flich},
  \& {Straniero}}]{Dominguez01}
{Dom{\'{\i}}nguez}, I., {H{\"o}flich}, P., \& {Straniero}, O. 2001, \apj, 557,
  279

\bibitem[{{Filippenko}(1997)}]{Filippenko97}
{Filippenko}, A.~V. 1997, \araa, 35, 309

\bibitem[{{Filippenko} {et~al.}(1992)}]{Filippenko92b}
{Filippenko}, A.~V. {et~al.} 1992, \aj, 104, 1543

\bibitem[{{Fink} {et~al.}(2010)}]{Fink10}
{Fink}, M. {et~al.} 2010, ArXiv e-prints

\bibitem[{{Gallagher} {et~al.}(2008)}]{Gallagher08}
{Gallagher}, J.~S. {et~al.} 2008, \apj, 685, 752

\bibitem[{{Ganeshalingam} {et~al.}(2011){Ganeshalingam}, {Li}, \&
  {Filippenko}}]{Ganeshalingam11}
{Ganeshalingam}, M., {Li}, W., \& {Filippenko}, A.~V. 2011, \mnras, 1265

\bibitem[{{Garnavich} {et~al.}(2004)}]{Garnavich04}
{Garnavich}, P.~M. {et~al.} 2004, \apj, 613, 1120

\bibitem[{{Goldhaber} {et~al.}(2001)}]{Goldhaber01}
{Goldhaber}, G. {et~al.} 2001, \apj, 558, 359

\bibitem[{{Gonz{\'a}lez-Gait{\'a}n} {et~al.}(2011)}]{Gonzalez11}
{Gonz{\'a}lez-Gait{\'a}n}, S. {et~al.} 2011, \apj, 727, 107

\bibitem[{{Guy} {et~al.}(2007)}]{Guy07}
{Guy}, J. {et~al.} 2007, \aap, 466, 11

\bibitem[{{Guy} {et~al.}(2010)}]{Guy10}
---. 2010, \aap, 523, A7+

\bibitem[{{Hamuy} {et~al.}(1996){Hamuy}, {Phillips}, {Suntzeff}, {Schommer},
  {Maza}, \& {Aviles}}]{Hamuy96}
{Hamuy}, M., {Phillips}, M.~M., {Suntzeff}, N.~B., {Schommer}, R.~A., {Maza},
  J., \& {Aviles}, R. 1996, \aj, 112, 2391

\bibitem[{{Hayden} {et~al.}(2010{\natexlab{a}})}]{Hayden10b}
{Hayden}, B.~T. {et~al.} 2010{\natexlab{a}}, ArXiv e-prints

\bibitem[{{Hayden} {et~al.}(2010{\natexlab{b}})}]{Hayden10}
---. 2010{\natexlab{b}}, ArXiv:1001.3428

\bibitem[{{Hicken} {et~al.}(2007)}]{Hicken07}
{Hicken}, M. {et~al.} 2007, \apj, 669, L17

\bibitem[{{Hicken} {et~al.}(2009)}]{Hicken09a}
---. 2009, \apj, 700, 331

\bibitem[{{Hoeflich} {et~al.}(1993){Hoeflich}, {Mueller}, \&
  {Khokhlov}}]{Hoeflich93}
{Hoeflich}, P., {Mueller}, E., \& {Khokhlov}, A. 1993, \aap, 268, 570

\bibitem[{{H{\"o}flich} {et~al.}(2002)}]{Hoeflich02}
{H{\"o}flich}, P. {et~al.} 2002, \apj, 568, 791

\bibitem[{{H{\"o}flich} {et~al.}(2010)}]{Hoeflich10}
---. 2010, \apj, 710, 444

\bibitem[{{Holtzman} {et~al.}(2008)}]{Holtzman08}
{Holtzman}, J.~A. {et~al.} 2008, \aj, 136, 2306

\bibitem[{{Howell}(2001)}]{Howell01a}
{Howell}, D.~A. 2001, \apjl, 554, L193

\bibitem[{{Howell} {et~al.}(2009)}]{Howell09}
{Howell}, D.~A. {et~al.} 2009, \apj, 691, 661

\bibitem[{{Hsiao} {et~al.}(2007){Hsiao}, {Conley}, {Howell}, {Sullivan},
  {Pritchet}, {Carlberg}, {Nugent}, \& {Phillips}}]{Hsiao07}
{Hsiao}, E.~Y., {Conley}, A., {Howell}, D.~A., {Sullivan}, M., {Pritchet},
  C.~J., {Carlberg}, R.~G., {Nugent}, P.~E., \& {Phillips}, M.~M. 2007, \apj,
  663, 1187

\bibitem[{{Jha} {et~al.}(2007){Jha}, {Riess}, \& {Kirshner}}]{Jha07}
{Jha}, S., {Riess}, A.~G., \& {Kirshner}, R.~P. 2007, \apj, 659, 122

\bibitem[{{Jha} {et~al.}(2006)}]{Jha06}
{Jha}, S. {et~al.} 2006, \aj, 131, 527

\bibitem[{{Kasen}(2010)}]{Kasen10}
{Kasen}, D. 2010, \apj, 708, 1025

\bibitem[{{Kasen} {et~al.}(2009){Kasen}, {R{\"o}pke}, \& {Woosley}}]{Kasen09}
{Kasen}, D., {R{\"o}pke}, F.~K., \& {Woosley}, S.~E. 2009, \nat, 460, 869

\bibitem[{{Kasliwal} {et~al.}(2010)}]{Kasliwal10}
{Kasliwal}, M.~M. {et~al.} 2010, \apjl, 723, L98

\bibitem[{{Kelly} {et~al.}(2010)}]{Kelly10}
{Kelly}, P.~L. {et~al.} 2010, \apj, 715, 743

\bibitem[{{Khokhlov} {et~al.}(1993){Khokhlov}, {Mueller}, \&
  {Hoeflich}}]{Khokhlov93}
{Khokhlov}, A., {Mueller}, E., \& {Hoeflich}, P. 1993, \aap, 270, 223

\bibitem[{{Khokhlov}(1991)}]{Khokhlov91}
{Khokhlov}, A.~M. 1991, \aap, 245, 114

\bibitem[{{Krisciunas} {et~al.}(2003)}]{Krisciunas03}
{Krisciunas}, K. {et~al.} 2003, \aj, 125, 166

\bibitem[{{Krisciunas} {et~al.}(2004)}]{Krisciunas04}
---. 2004, \aj, 127, 1664

\bibitem[{{Leibundgut} {et~al.}(1993)}]{Leibundgut93}
{Leibundgut}, B. {et~al.} 1993, \aj, 105, 301

\bibitem[{{Leonard} {et~al.}(2005){Leonard}, {Li}, {Filippenko}, {Foley}, \&
  {Chornock}}]{Leonard05}
{Leonard}, D.~C., {Li}, W., {Filippenko}, A.~V., {Foley}, R.~J., \& {Chornock},
  R. 2005, \apj, 632, 450

\bibitem[{{Lira} {et~al.}(1998)}]{Lira98}
{Lira}, P. {et~al.} 1998, \aj, 115, 234

\bibitem[{{Livne}(1990)}]{Livne90}
{Livne}, E. 1990, \apj, 354, L53

\bibitem[{{Maeda} {et~al.}(2010)}]{Maeda10}
{Maeda}, K. {et~al.} 2010, \apj, 712, 624

\bibitem[{{Mazzali} {et~al.}(2007){Mazzali}, {R{\"o}pke}, {Benetti}, \&
  {Hillebrandt}}]{Mazzali07}
{Mazzali}, P.~A., {R{\"o}pke}, F.~K., {Benetti}, S., \& {Hillebrandt}, W. 2007,
  Science, 315, 825

\bibitem[{{McClelland} {et~al.}(2010)}]{McClelland10}
{McClelland}, C.~M. {et~al.} 2010, \apj, 720, 704

\bibitem[{{Neill} {et~al.}(2011)}]{Neill11}
{Neill}, J.~D. {et~al.} 2011, \apj, 727, 15

\bibitem[{{Nugent} {et~al.}(1995){Nugent}, {Branch}, {Baron}, {Fisher},
  {Vaughan}, \& {Hauschildt}}]{Nugent95}
{Nugent}, P., {Branch}, D., {Baron}, E., {Fisher}, A., {Vaughan}, T., \&
  {Hauschildt}, P.~H. 1995, Physical Review Letters, 75, 394

\bibitem[{{Nugent} {et~al.}(2002){Nugent}, {Kim}, \& {Perlmutter}}]{Nugent02}
{Nugent}, P., {Kim}, A., \& {Perlmutter}, S. 2002, \pasp, 114, 803

\bibitem[{{Pakmor} {et~al.}(2010){Pakmor}, {Kromer}, {R{\"o}pke}, {Sim},
  {Ruiter}, \& {Hillebrandt}}]{Pakmor10}
{Pakmor}, R., {Kromer}, M., {R{\"o}pke}, F.~K., {Sim}, S.~A., {Ruiter}, A.~J.,
  \& {Hillebrandt}, W. 2010, \nat, 463, 61

\bibitem[{{Pastorello} {et~al.}(2007)}]{Pastorello07}
{Pastorello}, A. {et~al.} 2007, \mnras, 376, 1301

\bibitem[{{Perrett} {et~al.}(2011)}]{Perrett11}
{Perrett}, K. {et~al.} 2011, in preparation

\bibitem[{{Phillips}(1993)}]{Phillips93}
{Phillips}, M.~M. 1993, \apjl, 413, L105

\bibitem[{{Phillips} {et~al.}(1999){Phillips}, {Lira}, {Suntzeff}, {Schommer},
  {Hamuy}, \& {Maza}}]{Phillips99}
{Phillips}, M.~M., {Lira}, P., {Suntzeff}, N.~B., {Schommer}, R.~A., {Hamuy},
  M., \& {Maza}, J.~. 1999, \aj, 118, 1766

\bibitem[{{Phillips} {et~al.}(1992)}]{Phillips92}
{Phillips}, M.~M. {et~al.} 1992, \aj, 103, 1632

\bibitem[{{Poznanski} {et~al.}(2010)}]{Poznanski10}
{Poznanski}, D. {et~al.} 2010, Science, 327, 58

\bibitem[{{Riess} {et~al.}(1999)}]{Riess99b}
{Riess}, A.~G. {et~al.} 1999, \aj, 118, 2675

\bibitem[{{Riess} {et~al.}(2005)}]{Riess05}
---. 2005, \apj, 627, 579

\bibitem[{{Roepke} {et~al.}(2010)}]{Roepke10}
{Roepke}, F.~K. {et~al.} 2010, ArXiv e-prints

\bibitem[{{Schwarz}(1978)}]{Schwarz78}
{Schwarz}, G.~E. 1978, Annals of Statistics, 6, 461

\bibitem[{{Shigeyama} {et~al.}(1992)}]{Shigeyama92}
{Shigeyama}, T. {et~al.} 1992, \apjl, 386, L13

\bibitem[{{Sim} {et~al.}(2010){Sim}, {R{\"o}pke}, {Hillebrandt}, {Kromer},
  {Pakmor}, {Fink}, {Ruiter}, \& {Seitenzahl}}]{Sim10}
{Sim}, S.~A., {R{\"o}pke}, F.~K., {Hillebrandt}, W., {Kromer}, M., {Pakmor},
  R., {Fink}, M., {Ruiter}, A.~J., \& {Seitenzahl}, I.~R. 2010, \apjl, 714, L52

\bibitem[{{Smith} {et~al.}(2002)}]{Smith02}
{Smith}, J.~A. {et~al.} 2002, \aj, 123, 2121

\bibitem[{{Stanishev} {et~al.}(2007)}]{Stanishev07}
{Stanishev}, V. {et~al.} 2007, \aap, 469, 45

\bibitem[{{Strovink}(2007)}]{Strovink07}
{Strovink}, M. 2007, \apj, 671, 1084

\bibitem[{{Sullivan} {et~al.}(2010)}]{Sullivan10}
{Sullivan}, M. {et~al.} 2010, \mnras, 406, 782

\bibitem[{{Taubenberger} {et~al.}(2008)}]{Taubenberger08}
{Taubenberger}, S. {et~al.} 2008, \mnras, 385, 75

\bibitem[{{Taylor}(1997)}]{Taylor97}
{Taylor}, J. 1997, {Introduction to Error Analysis, the Study of Uncertainties
  in Physical Measurements, 2nd Edition}, ed. {Taylor, J.} (University Science
  Books)

\bibitem[{{Timmes} {et~al.}(2003){Timmes}, {Brown}, \& {Truran}}]{Timmes03}
{Timmes}, F.~X., {Brown}, E.~F., \& {Truran}, J.~W. 2003, \apjl, 590, L83

\bibitem[{{Truran} {et~al.}(1967){Truran}, {Arnett}, \& {Cameron}}]{Truran67}
{Truran}, J.~W., {Arnett}, W.~D., \& {Cameron}, A.~G.~W. 1967, Canadian Journal
  of Physics, 45, 2315

\bibitem[{{Tsvetkov}(2006{\natexlab{a}})}]{Tsvetkov06a}
{Tsvetkov}, D.~Y. 2006{\natexlab{a}}, Peremennye Zvezdy, 26, 3

\bibitem[{{Tsvetkov}(2006{\natexlab{b}})}]{Tsvetkov06c}
---. 2006{\natexlab{b}}, Peremennye Zvezdy, 26, 4

\bibitem[{{Valentini} {et~al.}(2003)}]{Valentini03}
{Valentini}, G. {et~al.} 2003, \apj, 595, 779

\bibitem[{{van Kerkwijk} {et~al.}(2010){van Kerkwijk}, {Chang}, \&
  {Justham}}]{vanKerkwijk10}
{van Kerkwijk}, M.~H., {Chang}, P., \& {Justham}, S. 2010, \apjl, 722, L157

\bibitem[{{Wang} {et~al.}(2009)}]{Wang09}
{Wang}, X. {et~al.} 2009, \apj, 697, 380

\bibitem[{{Woosley} \& {Weaver}(1994)}]{Woosley94}
{Woosley}, S.~E. \& {Weaver}, T.~A. 1994, \apj, 423, 371

\bibitem[{{Yamaoka} {et~al.}(1992)}]{Yamaoka92}
{Yamaoka}, H. {et~al.} 1992, \apjl, 393, L55

\end{thebibliography}

\end{document}